\documentclass[reprint, amsmath,amssymb, superscriptaddress, aps, prc,showpacs, showkeys,nofootinbib, nobibnotes]{revtex4-1}
\usepackage{graphicx}
\usepackage{dcolumn}
\usepackage{bm}
\usepackage{ulem} 
\usepackage{cancel}
\usepackage{hyperref}
\usepackage[mathlines]{lineno}
\usepackage{natbib}
\usepackage{titlesec}
\usepackage{float}

\DeclareSymbolFont{extraup}{U}{zavm}{m}{n}
\DeclareMathSymbol{\varheart}{\mathalpha}{extraup}{86}
\DeclareMathSymbol{\vardiamond}{\mathalpha}{extraup}{87}

\begin{document}

\title{Structure of $^{83}$As, $^{85}$As and $^{87}$As: from semi-magicity to $\gamma$-softness}
\author{K.~Rezynkina}
\email{kseniia.rezynkina@pd.infn.it}
\affiliation{Université de Strasbourg, CNRS, IPHC UMR 7178, F-67000 Strasbourg, France}
\affiliation{INFN Sezione di Padova, I-35131 Padova, Italy}
\author{D. D.~Dao}
\affiliation{Université de Strasbourg, CNRS, IPHC UMR 7178, F-67000 Strasbourg, France}
\author{G.~Duch\^ene}
\affiliation{Université de Strasbourg, CNRS, IPHC UMR 7178, F-67000 Strasbourg, France}
\author{J.~Dudouet}
\affiliation{Université Claude Bernard Lyon, CNRS/IN2P3, IP2I Lyon, F-69622, Villeurbanne, France}
\author{F.~Nowacki}
\affiliation{Université de Strasbourg, CNRS, IPHC UMR 7178, F-67000 Strasbourg, France}
\author{E.~Clément}
\affiliation{GANIL, CEA/DRF-CNRS/IN2P3, BP 55027, 14076 Caen Cedex 5, France}
\author{A.~Lemasson}
\affiliation{GANIL, CEA/DRF-CNRS/IN2P3, BP 55027, 14076 Caen Cedex 5, France}
\author{C.~Andreoiu}
\affiliation{Department of Chemistry, Simon Fraser University, Burnaby, British Columbia V5A 1S6, Canada}
\author{A.~Astier}
\affiliation{Université Paris-Saclay, CNRS/IN2P3, IJCLab, 91405 Orsay, France}
\author{G.~de~Angelis}
\affiliation{INFN, Laboratori Nazionali di Legnaro, Viale dell'Universtità 2, I-35020 Legnaro (PD), Italy}
\author{G.~de~France}
\affiliation{GANIL, CEA/DRF-CNRS/IN2P3, BP 55027, 14076 Caen Cedex 5, France}
\author{C.~Delafosse}
\affiliation{Université Paris-Saclay, CNRS/IN2P3, IJCLab, 91405 Orsay, France}
\author{I.~Deloncle}
\affiliation{Université Paris-Saclay, CNRS/IN2P3, IJCLab, 91405 Orsay, France}
\author{F.~Didierjean}
\affiliation{Université de Strasbourg, CNRS, IPHC UMR 7178, F-67000 Strasbourg, France}
\author{Z.~Dombradi}
\affiliation{Institute for Nuclear Research of the Hungarian Academy of Sciences, Pf.51, H-4001, Debrecen, Hungary}
\author{C.~Ducoin}
\affiliation{Université Lyon, Université Claude Bernard Lyon, CNRS/IN2P3, IP2I Lyon, F-69622, Villeurbanne, France}
\author{A.~Gadea}
\affiliation{Instituto de Física Corpuscular, CSIC-Universitat de València, E-46980 Valencia, Spain}
\author{A.~Gottardo}
\affiliation{Université Paris-Saclay, CNRS/IN2P3, IJCLab, 91405 Orsay, France}
\affiliation{INFN, Laboratori Nazionali di Legnaro, Viale dell'Universtità 2, I-35020 Legnaro (PD), Italy}
\author{D.~Guinet}
\affiliation{Université Lyon, Université Claude Bernard Lyon, CNRS/IN2P3, IP2I Lyon, F-69622, Villeurbanne, France}
\author{B.~Jacquot}
\affiliation{GANIL, CEA/DRF-CNRS/IN2P3, BP 55027, 14076 Caen Cedex 5, France}
\author{P.~Jones}
\affiliation{iThemba LABS, National Research Foundation, P.O.Box 722, Somerset West,7129 South Africa}
\author{T.~Konstantinopoulos}
\affiliation{Université Paris-Saclay, CNRS/IN2P3, IJCLab, 91405 Orsay, France}
\author{I.~Kuti}
\affiliation{Institute for Nuclear Research of the Hungarian Academy of Sciences, Pf.51, H-4001, Debrecen, Hungary}
\author{A.~Korichi}
\affiliation{Université Paris-Saclay, CNRS/IN2P3, IJCLab, 91405 Orsay, France}
\author{S.M.~Lenzi}
\affiliation{INFN Sezione di Padova, I-35131 Padova, Italy}
\affiliation{Dipartimento di Fisica e Astronomia dell’Università di Padova, I-35131 Padova, Italy}
\author{G.~Li}
\affiliation{GSI, Helmholtzzentrum für Schwerionenforschung GmbH, D-64291 Darmstadt, Germany}
\affiliation{Institute of Modern Physics, Chinese Academy of Sciences, Lanzhou, 73000, China}
\author{F. Le Blanc}
\affiliation{Université de Strasbourg, CNRS, IPHC UMR 7178, F-67000 Strasbourg, France}
\affiliation{Université Paris-Saclay, CNRS/IN2P3, IJCLab, 91405 Orsay, France}
\author{C. Lizarazo}
\affiliation{Institut für Kernphysik, Technische Universität Darmstadt, D-64289 Darmstadt, Germany}
\author{R.~Lozeva}
\affiliation{Université de Strasbourg, CNRS, IPHC UMR 7178, F-67000 Strasbourg, France}
\affiliation{Université Paris-Saclay, CNRS/IN2P3, IJCLab, 91405 Orsay, France}
\author{G.~Maquart}
\affiliation{Université Lyon, Université Claude Bernard Lyon, CNRS/IN2P3, IP2I Lyon, F-69622, Villeurbanne, France}
\author{B.~Million}
\affiliation{INFN, Sezione di Milano, I-20133 Milano, Italy}
\author{C. Michelagnoli}
\affiliation{GANIL, CEA/DRF-CNRS/IN2P3, BP 55027, 14076 Caen Cedex 5, France}
\affiliation{Institut Laue-Langevin, 71 avenue des Martyrs, 38042 Grenoble Cedex 9, France}
\author{D.R.~Napoli}
\affiliation{INFN, Laboratori Nazionali di Legnaro, Viale dell'Universtità 2, I-35020 Legnaro (PD), Italy}
\author{A.~Navin}
\affiliation{GANIL, CEA/DRF-CNRS/IN2P3, BP 55027, 14076 Caen Cedex 5, France}
\author{R.M.~Pérez-Vidal}
\affiliation{Instituto de Física Corpuscular, CSIC-Universitat de València, E-46980 Valencia, Spain}
\author{C.M.~Petrache}
\affiliation{Université Paris-Saclay, CNRS/IN2P3, IJCLab, 91405 Orsay, France}
\author{N. Pietralla}
\affiliation{Institut für Kernphysik, Technische Universität Darmstadt, D-64289 Darmstadt, Germany}
\author{D.~Ralet}
\affiliation{Institut für Kernphysik, Technische Universität Darmstadt, D-64289 Darmstadt, Germany}
\affiliation{MIRION Technologies Canberra, 1 Chemin de la Roseraie, 67380 Lingolsheim, France}
\author{M.~Ramdhane}
\affiliation{LPSC, Université Grenoble-Alpes, CNRS/IN2P3, 38026 Grenoble Cedex, France}
\author{M. Rejmund}
\affiliation{GANIL, CEA/DRF-CNRS/IN2P3, BP 55027, 14076 Caen Cedex 5, France}
\author{O. Stezowski}
\affiliation{Université Lyon, Université Claude Bernard Lyon, CNRS/IN2P3, IP2I Lyon, F-69622, Villeurbanne, France}
\author{C.~Schmitt}
\affiliation{GANIL, CEA/DRF-CNRS/IN2P3, BP 55027, 14076 Caen Cedex 5, France}
\affiliation{Université de Strasbourg, CNRS, IPHC UMR 7178, F-67000 Strasbourg, France}
\author{D.~Sohler}
\affiliation{Institute for Nuclear Research of the Hungarian Academy of Sciences, Pf.51, H-4001, Debrecen, Hungary}
\author{D.~Verney}
\affiliation{Université Paris-Saclay, CNRS/IN2P3, IJCLab, 91405 Orsay, France}

\date{\today}

\begin{abstract}
The structure of $^{83}$As, $^{85}$As and $^{87}$As have been studied in fusion-fission reaction $^{238}$U+$^9$Be. Fission fragments were identified in mass and atomic number using the VAMOS++ spectrometer and the coincident $\gamma$-rays were detected in the $\gamma$-ray tracking array AGATA. New transitions in $^{83}$As and $^{85}$As are reported and placed in the level schemes.  A level scheme of the excited states in $^{87}$As is proposed for the first time. The data are interpreted in frame of Large-Scale Shell-Model calculations, SU3 symmetries and Beyond Mean-Field frameworks. A spherical regime at magic number $N$=50 is predicted and the location of the proton $g_{9/2}$ orbital is proposed for the first time. Development of collectivity in a prolate deformed, $\gamma$-soft regime in the open shell cases $^{85}$As and $^{87}$As, most neutron-rich isotopes beyond $N$=50, is concluded. Data and theoretical calculations give confidence to a relatively high extrapolated excitation energy about 4 MeV of the $9/2^+$ state in $^{79}$Cu, one proton above $^{78}$Ni.\\

\end{abstract}


\maketitle
\section{\label{sec:intr}Introduction}

The region of doubly-magic $^{78}$Ni and structure evolution in its vicinity is rich and challenging for the nuclear spectroscopy studies. Recent experimental observations demonstrated that $^{78}$Ni remains a doubly-magic nucleus,  while a prolate shape structure built on top of the $0_2^+$ state has been observed to appear at rather low excitation energy \cite{78Ni, Nowacki}.
The interplay of single-particle and multi-nucleon excitations near the two closed shells gives rise to collective effects, which attracts much interest in the nuclear structure community. Beyond $N$=50, in $^{84}$Ge, $^{86}$Ge, $^{88}$Ge the occurrence of triaxial deformation has been predicted by both Large Scale Shell Model (LSSM) and GCM+GOA  calculations, and found experimentally \cite{Verney,Lettmann}. In $^{80}$Ge the possibility of occurrence of shape coexistence has been explored both in experimental works \cite{Gottardo80Ge, Triumf80Ge, ISOLDE80Ge} and in theoretical studies (see e.g. \cite{Theor80GeYES}).

In the selenium chain, experimental findings as well as LSSM and Beyond  Mean-field interpretations \cite{LSSM-BMF,Se86,Se-Riken} point also to large collectivity. Therefore arsenic isotopes lie just in the transition region, and can be seen as a particle coupled to a germanium core, or as a hole coupled to a selenium core.\\
The accurate understanding of the underlying structure of nuclei beyond $Z = 28$ and $N = 50$ is critical for a broad range of research directions including nuclear energy and \textit{r-}process nucleosynthesis. 
Proper definition of shell gaps and description of the onset of collectivity are fundamental to gain predictive power in the entire region surrounding $^{78}$Ni. Yet, the available experimental data are, to date, very scarce. Besides, description and interpretation of various excitation types are very challenging since these nuclei lie in the transition region from harmonic oscillator to spin-orbit shell closures ($Z$=$N$=40 is no longer a strong shell closure) but where core excitations are  important as recently seen in $^{78}$Ni \cite{78Ni} and $^{81}$Ga \cite{Dud81Ga}.\\
In this work we aim to extend the knowledge of this region of the nuclear chart. We focus on the odd-mass arsenic isotopic chain to study the evolution of the nuclear structure between $^{83}$As, with a closed neutron shell, towards $^{87}$As, with 4 neutrons above the $N$=50 shell gap.\par
As mentioned above,  deformation regime has been firmly established experimentally in the germanium and selenium isotopes with $N$=52-56 \cite{Verney,Lettmann} and therefore arsenic isotopes with the same neutron number are expected to exhibit collective regime, with the exception of semi-magic $^{83}$As.
\begin{figure*}[t]
\centering
\includegraphics[width=1\linewidth]{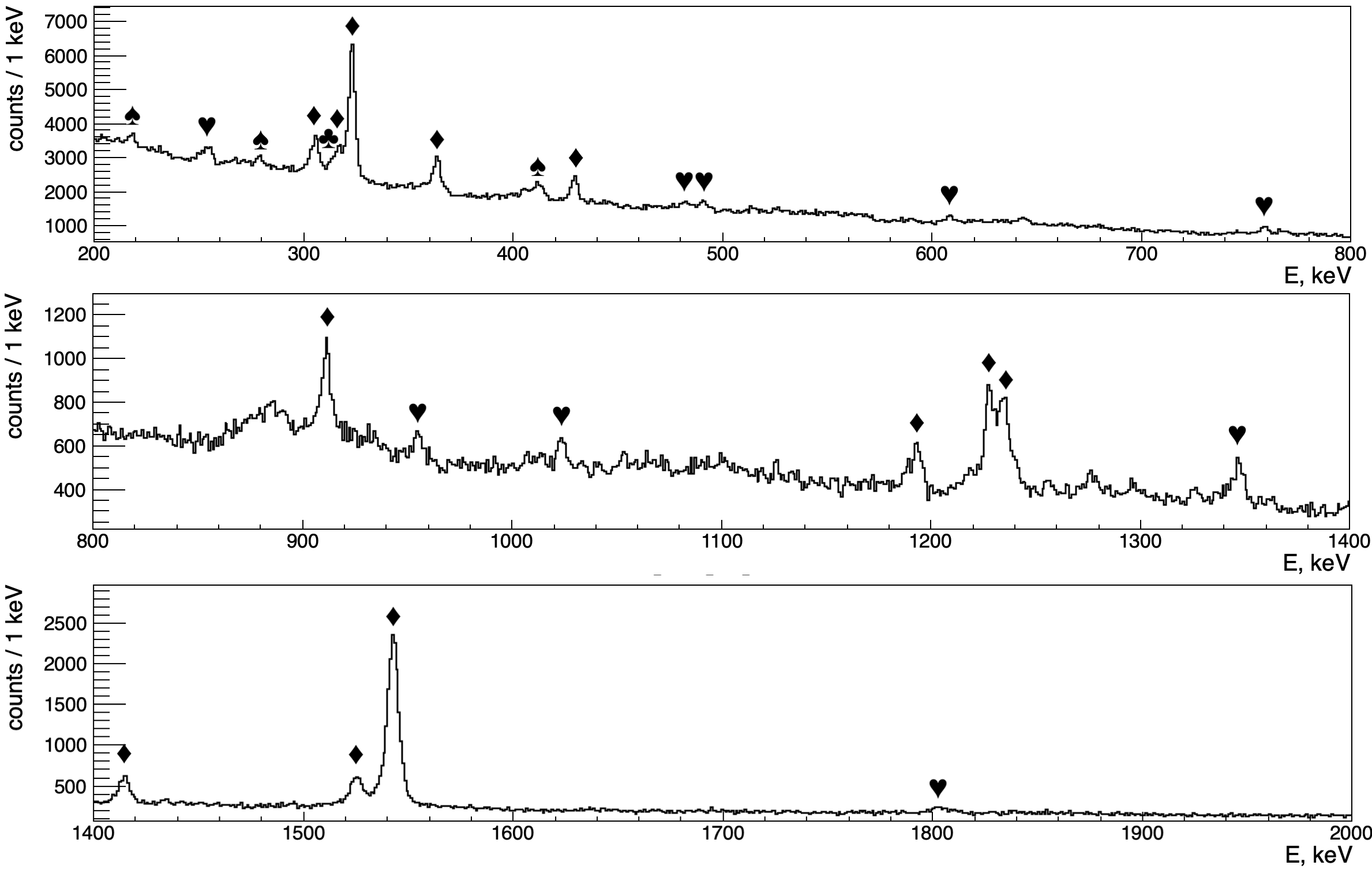}
\caption{Tracked $\gamma$-ray spectrum in coincidence with $^{83}$As ions identified in VAMOS++.  ($\varheart$): $^{83}$As transitions reported for the first time in this work; ($\vardiamond$): $^{83}$As transitions previously observed in \cite{Porquet,Sahin,Drouet,Baczyk}; ($\spadesuit$):  $^{82}$As contaminant transitions; ($\clubsuit$) $^{84}$As contaminant transitions.}
\label{83As_singles}
\end{figure*}

\section{\label{sec:ExpDet} Experimental details}
The exotic $^{83,85,87}$As nuclei were produced in GANIL, in a fusion-fission reaction with a $^{238}$U beam at 6.2~MeV/u energy and 1~pnA average intensity impinging upon a 10~$\mu$m thick $^9$Be target. The prompt $\gamma$-rays depopulating the medium-spin excited states were measured with the AGATA $\gamma$-ray tracking array~\cite{AGATA, Korichi}. The identification of the nuclei was performed on an event-by-event basis with the VAMOS++ spectrometer~\cite{VAMOS1, VAMOS2}, which provided the determination of mass number ($A$) and proton number ($Z$) of the incoming fission fragments, as well as their velocity. VAMOS++ was placed at 28$^{\circ}$ with respect to the beam direction, in order to maximize its acceptance for the neutron-rich low-$Z$ fission fragments. The AGATA spectrometer consisted of 8 triple clusters of segmented HPGe crystals \cite{Clement} placed around the target. The Doppler correction was performed on an event-by-event basis.
The tracked $\gamma$-ray detection efficiency was $\sim$1.6\% at 500~keV and $\sim$1.2\% at 1~MeV. The interaction points of $\gamma$-rays in the sensitive volume of the detection array were determined with the adaptive grid search method~\cite{Venturelli}, and $\gamma$-ray tracking~\cite{track} was used for the reconstruction of the scattering pattern of each $\gamma$-ray. More details about the experimental setup can be found in \cite{Clement}.\par

In total, 205 species of nuclei were identified in AGATA-VAMOS++ coincidence events, with a wealth of new data which already allowed to study the structure of neutron-rich isotopes $^{81}$Ga \cite{Dud81Ga} and $^{96}$Kr \cite{Dud96Kr} from the same data set. Analyses for several other exotic isotopes are currently underway.\par

The asymmetric configuration of AGATA detectors in this experiment, made as compact as possible in order to enhance the detection efficiency, as well as the limited acquired $\gamma$-ray statistics for the nuclei under study, do not allow to establish spins and parities of the states based on angular distributions. Thus, most spin assignments in this work are made under the assumption of the enhanced probability to populate yrast medium- and high-spin states in fusion-fission reaction \cite{Navin}. The results of LSSM calculations (see Section \ref{sec:Discussion} for detail) were also used to guide the experimental spin assignments.\par

In sections \ref{sec:83As}, \ref{sec:85As} and \ref{sec:87As} experimental findings for $^{83}$As, $^{85}$As and $^{87}$As isotopes, respectively, are presented. The interpretation of these results and comparison with the predictions from Large-Scale Shell-Model (LSSM) and Beyond Mean-Field (BMF) calculations are discussed in section \ref{sec:Discussion}.\\

\section{\label{sec:83As}$^{83}$As}

\begin{table*}
\begin{center}
\begin{tabular}{c|c|c|c|c}
\hline
\hline
$E_{level},~keV $ & $J^\pi$ & $E_{\gamma}$,~keV & $I_{\gamma}$ & Coincidence \\
\hline
\hline
305.6 & $3/2^-$ & 305.6(1) & 12.4(7) & - \\	
\hline
1193.7 &  & 1193.7(2) & 5.8(7) & - \\
\hline
1414.5 &  & 1414.5(1) & 13.8(7) & - \\
\hline
1525.7 & ($7/2^-$) & 1525.7(1) & 16.4(7) & 608.4, 955.2 \\
\hline
1542.6 & (9/2$^-$) & 1542.6(1) & 100(1) & 317.2, 323.0, 363.6, 429.1,\\
					& & & & 482.4, 490.7, 758.7, 911.3, \\
					& & & & 1228.5, 1234.3, 1346.5, 1802.6 \\
\hline
1796.1 &  & 253.5(3) & 8(1) & 1542.6 \\
\hline
1865.6 & (11/2$^-$) & 323.0(1) & 54.3(8) & 317.2, 363.6, 911.3, 1228.5, \\
					& & & & 1542.6\\
\hline
2033.3 &  & 490.7(3) & 4.3(5) & 1542.6 \\
\hline
2134.1 & ($9/2^-$) & 608.4(3) & 3.5(5) & 1525.7 \\
	   &   & 2134.5(16) & 1.7(4) & - \\
\hline
2301.3 &  & 758.7(2) & 3.6(5) & 1542.6 \\
\hline
2348.0 &  & 482.4(4) & 3.3(5) & 323.0, 1542.6 \\
\hline
2480.9 & ($11/2^-$) & 955.2(3) & 2.9(4) & 1525.7, 2466 \\
\hline
2566.3 &  & 1023.7(3) & 2.7(4) & 1542.6 \\
\hline
2776.9 & ($11/2^-, 13/2^-$)  & 911.3(2)& 10.2(28) & 323.0, 1524.6 \\
&& 1234.3(2)&  14.9(6) & 1542.6 \\
\hline
2889.1 &  & 1346.5(2) & 6.4(6) & 1542.6 \\
\hline
3094.1 & (13/2$^-$) & 317.2(1) & 11.6(6) & 911.3, 1234.3, 1542.6 \\
 &  & 1228.5(1) & 15.7(7) & 323.0, 363.6,\\
\hline
3206.0 &  ($13/2^-, 15/2^-$)& 429.1(1) & 9.7(6) & 323.0, 1193.7, 1234.3, 1542.6,  2256\\
\hline
3345.2 &  & 1802.6(4) & 4.6(11) & 1542.6 \\
\hline
3457.7 & (15/2$^-$) & 363.6(1) & 15.5(8)& 323.0, 1228.5, 1542.6 \\
\hline
4947 & ($13/2^-, 15/2^-$) & 2466 & & 955.2 \\
\hline
5462 &  ($15/2^-$,$17/2^-$) & 2256 & & 429.1 \\
\hline
\hline
\end{tabular}
\caption{Relative $\gamma$-ray intensities observed in coincidence with the isotopically-identified $^{83}$As, normalized to the 1542.6~keV transition.}
\label{83As_tab}
\end{center}
\end{table*}

The excited states in $^{83}$As were previously studied in decay spectroscopy, via $\beta$-decay of $^{83}$Ge \cite{83GeBeta} and $\beta$-$n$ decay of $^{84}$Ge \cite{84GeBeta} as well as in-beam spectroscopy, in fusion-fission \cite{Porquet}, multi-nucleon transfer \cite{Sahin} and neutron-induced fission reactions \cite{Baczyk, Drouet}.

\begin{figure*}
\centering
\includegraphics[width=0.99\linewidth]{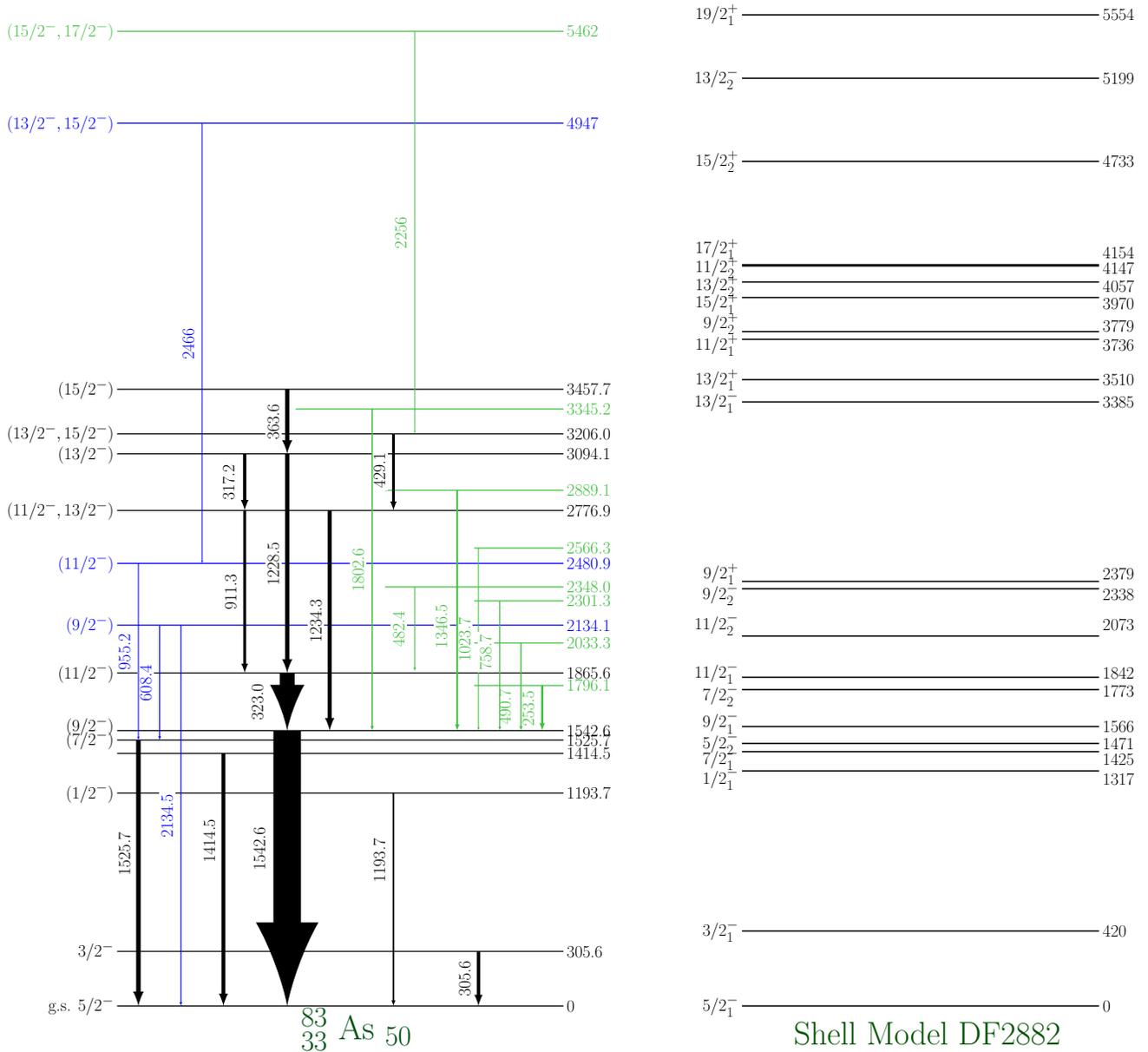} 
\caption{Left: Partial level scheme of $^{83}$As, containing the transitions observed in this work. The transitions indicated in color are new. The transitions marked in green eventually decay to the 1542.6~keV and 1865.6~keV levels. The transitions marked in blue follow other paths. Right: LSSM predictions with the DF2882 interaction. }
\label{83As_levels_theor}
\end{figure*}

The $\gamma$-ray spectrum observed in coincidence with the identification of $^{83}$As obtained in the present work is shown in Figure \ref{83As_singles}. The lines marked with a heart are observed for the first time in our work, while the other transitions (diamond) have been known from previous measurements. In total 1$\cdot 10^6$ $^{83}$As atoms were identified in the VAMOS++ spectrometer.
The complete list of $\gamma$-ray transitions observed in the present experiment, along with their intensities and with the coincident transitions, is given in Table \ref{83As_tab}. In Figure~\ref{83As_levels_theor} (left), a partial level scheme of $^{83}$As is presented, containing only the transitions observed in this work. The transitions marked in black have been observed in previous experiments~\cite{Porquet,Sahin,Drouet,Baczyk}. The news levels identified in this work are ordered and placed in the level scheme based on the observed $\gamma$-ray coincidences and on the relative intensities of the transitions.

A weak contamination from the neighbouring-mass arsenic isotopes $^{82}$As (the 218~keV, 279~keV and 411~keV transitions) and $^{84}$As (the 313~keV transition) is observed. It can be identified by comparing the spectra to ones gated on the corresponding isotopes identified in VAMOS++, or by tightening the mass selection. It is worth to remark that all the aforementioned transitions attributed to $^{82}$As and $^{84}$As are newly observed transitions.

A general agreement \cite{Porquet,Sahin,Drouet,Baczyk} is to assign $5/2^-$ to the ground state, even if shell-model calculations performed with different interactions predict close-lying $5/2^-$ and $3/2^-$ states.  A $3/2^-$ state at 305.6~keV excitation energy decays directly to the ground state, as proposed in Ref. \cite{Sahin} and confirmed in our work. The references \cite{Baczyk,Sahin,Drouet,Porquet} agree on the main gamma-ray cascade and spin-parity assignments which involve the 1542.6~keV, 323.0~keV,1228.5~keV and 363.6~keV transitions from the $5/2^-$ g.s.  up to the ($15/2^-$) state.  Parallel sequences are also observed with two pairs of transitions in cascade: 911.3~keV and 317.2~keV on one side and 1234.3~keV and 429.1~keV on the other. Spin assignments based on the literature and similarities with $^{85}$Br level scheme are proposed in \cite{Porquet}. Sahin et al. \cite{Sahin} suggest that the $9/2^+$ and $7/2^+$ states predicted by Shell Model calculations with the JJ4B interaction lie at about 2.5 and 3.4 MeV, respectively. In Baczyk et al. \cite{Baczyk} a transition at 1341.2~keV is reported for the first time, placed by the authors between the 3206.0~keV and 1865.6~keV levels.  In our data, we see no evidence of such transition.  In contrast, they did not observe the 317.2~keV transition connecting 3094.1~keV and 2776.9~keV levels.  Based on this, the positive-parity $9/2^+$, and ($11/2^+$, $13/2^+$) assignments were made in the aforementioned article to the 2776.9 and 3094.1~keV levels, respectively \cite{Baczyk}.\par

\begin{figure}
\centering
\includegraphics[width=0.99\linewidth]{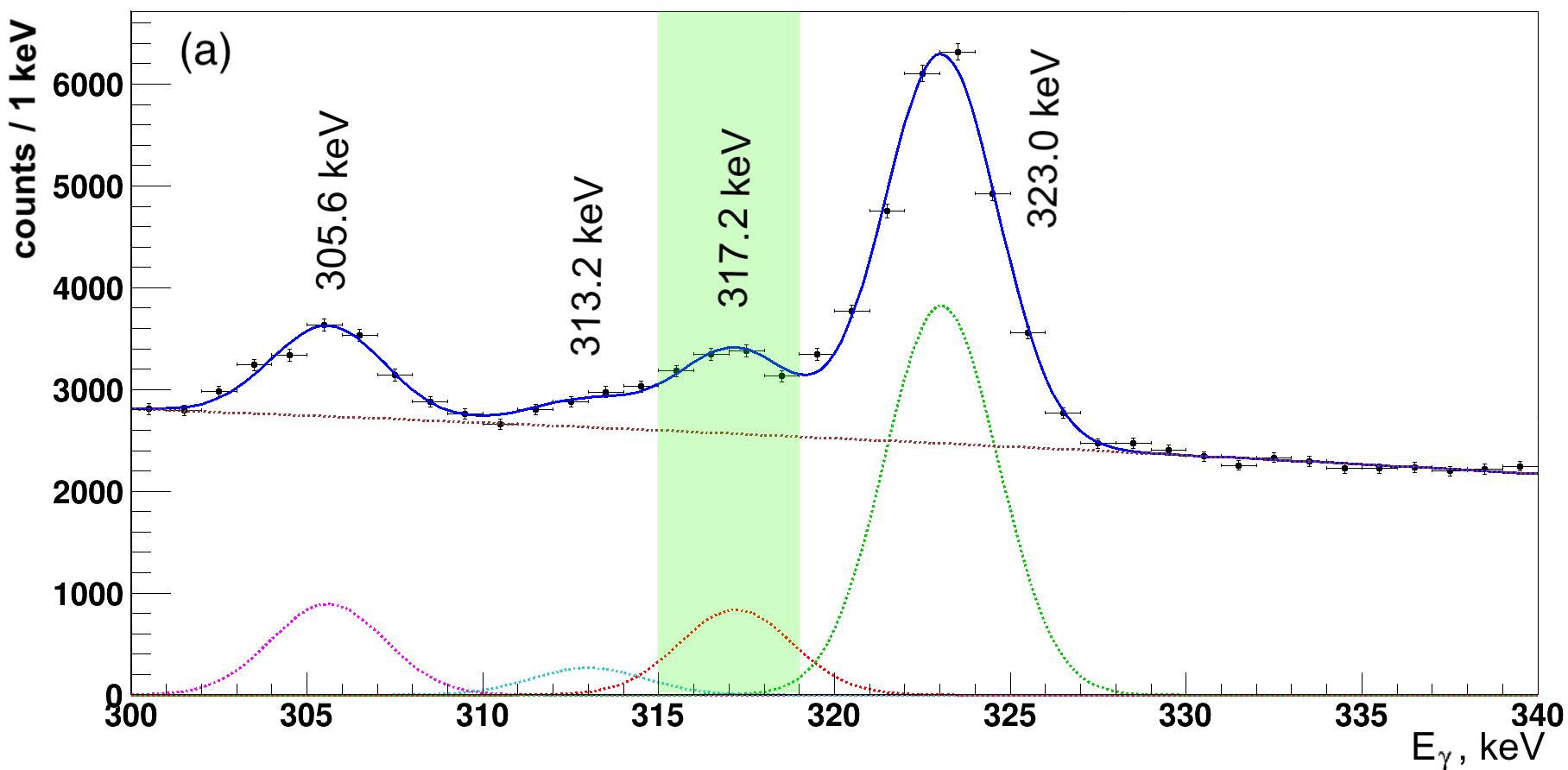} \\
\includegraphics[width=0.99\linewidth]{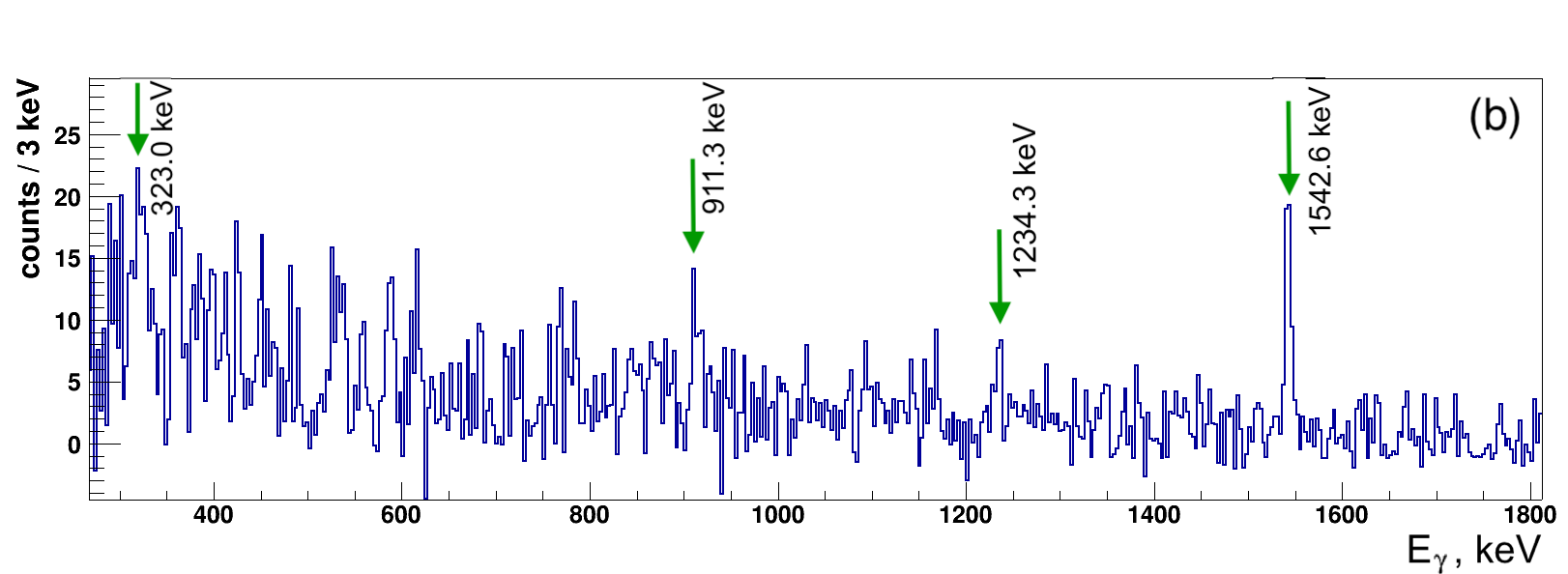} 
\caption{(a)Tracked $\gamma$-ray spectrum in coincidence with $^{83}$As ions identified in VAMOS++.  Zoom on the region near 320~keV.  The 317.2~keV transition is highlighted in green (color on-line). The weak 313.2(1)~keV line is a contaminant from $^{84}$As (see text for details); (b) background-subtracted gate on 317.2~keV transition in $^{83}$As (as highlighted in (a)).}
\label{83As_singles_zoom}
\end{figure}

Thanks to the high selectivity of VAMOS++ and efficiency of AGATA it was possible to observe, as illustrated in Figure~\ref{83As_singles_zoom} (a), the 317.2~keV line,  in contrast with the findings reported in Baczyk et al. \cite{Baczyk}.
This line is observed in coincidence with the 323.0, 363.6, 911.3, 1234.3 and 1542.6~keV transitions, as illustrated in Figure~\ref{83As_singles_zoom} (b), proving its correct placement (see Figure \ref{83As_levels_theor} (left)), and in agreement with the findings from Drouet et al.  \cite{Drouet}. 
While we do not observe the 1341~keV $\gamma$-ray line,  the 429.1~keV transition depopulating the same 3206.0~keV level is observed clearly with $\sim$15$\cdot 10^3$ counts and appears in coincidence with the 323.0, 911.3, 1234.3 and 1542.6~keV transitions. With the intensity ratio $I_{1341}:I_{429}=2:3$ reported in \cite{Baczyk}, after correcting for efficiency, we should observe about 1000 counts at 1341~keV, which is not the case. \par

\begin{figure}
\centering
\includegraphics[width=0.99\linewidth]{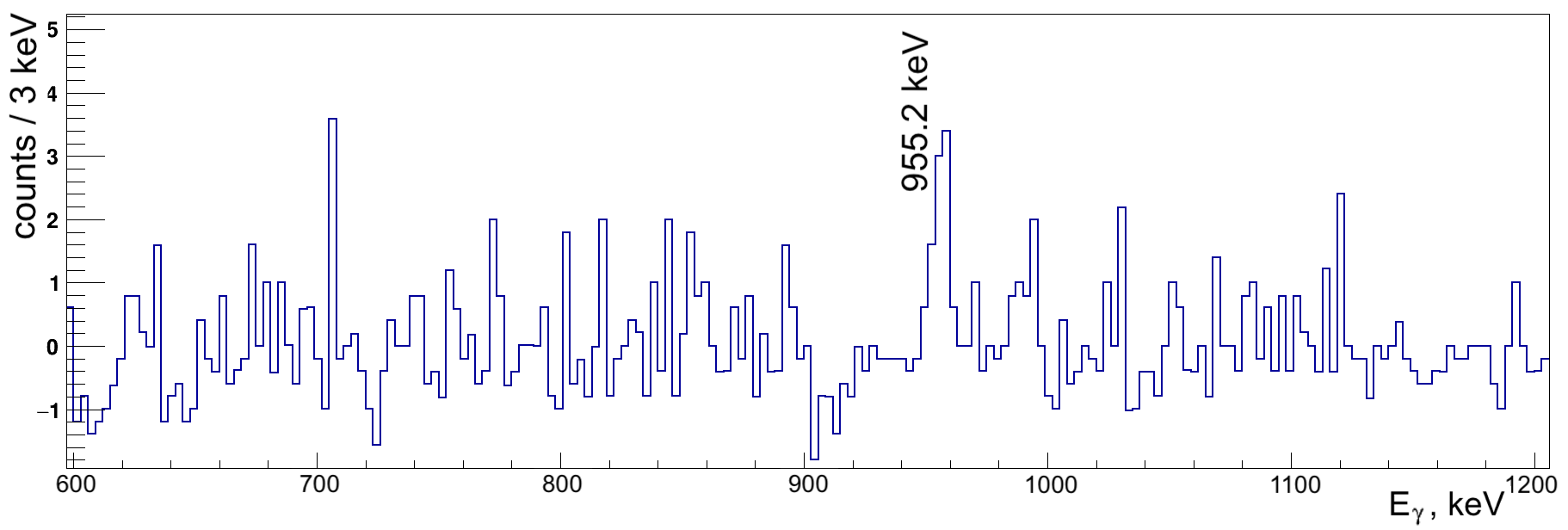} 
\caption{Background-subtracted  tracked $\gamma$-ray spectrum in coincidence with the $^{83}$As ions identified in VAMOS, gated on the 2466~keV $\gamma$-ray line.}
\label{83As_2466keV}
\end{figure}

Baczyk et al.~\cite{Baczyk} based the assignment of the $9/2^+$ spin and parity to the 2776.9 keV level on the fact that they haven't observed the 317.2~keV transition, and have observed the 1341.2~keV transition to this level. Having (in agreement with Drouet et al.~\cite{Drouet}) observed the reverse in our experiment: the 317.2~keV transition and not the 1341.2~keV one, we doubt this assignment. The multipolarity of the 1228.5~keV ($13/2^-$)$\rightarrow$($11/2^-$) transition was established through angular correlations, allowing to assign the spin and parity of the 3094.1~keV level \cite{Porquet}. Thus, if the 2776.9~keV level was a $9/2^+$, the 317.2~keV transition should have $M2$/$E3$ multipolarity. This would result in a partial lifetime of at least 700~ns for this excited state, which is more than 100 times longer than what could be observed in the present experiment. Indeed, the fast recoiling nuclei ($v/c\simeq0.1$) enter VAMOS++ placed about 30~cm downstream from the target, hindering the detection of $\gamma$-rays emitted $\leq$10~ns after the reaction. Moreover, such transition would not compete with the 1228.5~keV M1 decay from the $13/2^-$ state.  

\begin{figure*}
\centering
\includegraphics[width=0.99\linewidth]{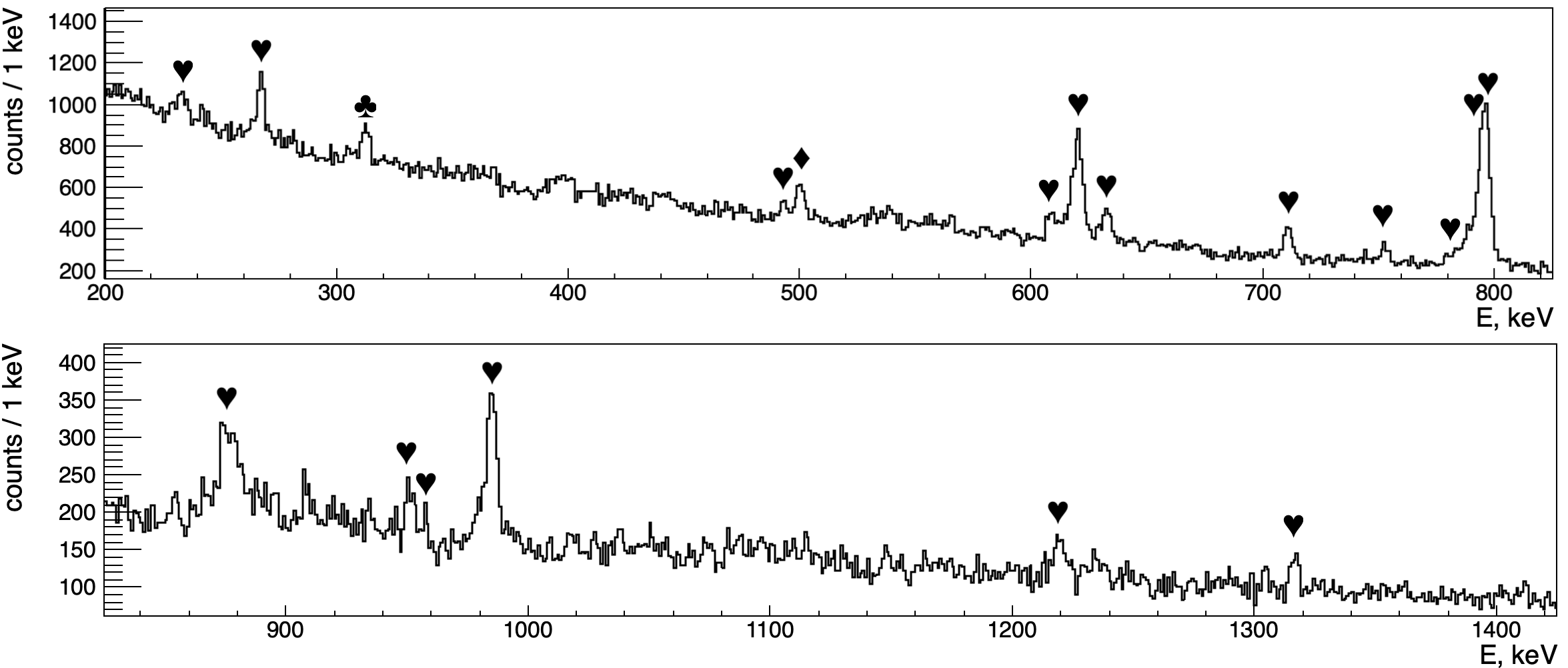} 
\caption{Tracked $\gamma$-ray spectrum in coincidence with $^{85}$As ions identified in VAMOS++.  ($\varheart$): $^{85}$As transitions reported for the first time in this work; ($\vardiamond$): $^{85}$As transitions previously observed in \cite{Korgul17}; ($\clubsuit$) $^{84}$As contaminant transitions.}
\label{85As_singles}
\end{figure*}

Based on the spin assignment of the main decay sequence with a ($13/2^-$) state decaying via 1228.5~keV transition to the ($11/2^-$) state at 1865.6~keV and via a cascade of two transitions 317.2~keV and 911.3~keV, we suggest that the 2776.9~keV state is assigned ($11/2^-$) or ($13/2^-$) spin and parity. The $9/2^+$ state may lie in an excitation energy range around 3 MeV as proposed by Baczyk et al. and could correspond to one of the non-assigned, newly placed states marked in green in Figure~\ref{83As_levels_theor}(left).
The multipolarity of the 429.1~keV transition being not established, we suggest the spin and parity ($13/2^-$) or ($15/2^-$) for the 3206.0~keV level.
The group of states traced in green in Figure~\ref{83As_levels_theor}(left) is observed in direct coincidence with the 1542.6~keV transition, or with the 323.0~keV transition feeding it.  No spin assignment was possible. The 955.2~keV and 608.4~keV transitions, traced in blue in Figure~\ref{83As_levels_theor}(left), are seen in coincidence with the 1525.7~keV $\gamma$-ray line, and not with each other.  The state decaying via the 955.2~keV transition to the  ($7/2^-$) state at 1525.7~keV can be tentatively assigned ($11/2^-$) spin and parity.  The 2134.5~keV transition is assigned to feed the ground state since it is not seen in coincidence with any other line and its energy corresponds within the uncertainty to the sum of 608.4~keV and 1525.7~keV transitions.  Therefore,  the state decaying to the ($7/2^-$) state at 1525.7~keV with the 608.4~keV transition and directly to the $5/2^-$ g.s. with the 2134.5~keV transition is likely to be ($9/2^-$).
As shown in Figure~\ref{83As_2466keV}, the high-energy 2466~keV transition is observed in coincidence with the 955.2~keV one, and thus de-excites a newly observed level at 4974~keV, the spin assignment of which may be ($11/2^-$,$13/2^-$) for a dipole transition and  ($13/2^-$,$15/2^-$) for a quadrupole transition.  As the fusion-fission reaction feeds preferentially yrast or near-yrast states, we suggest to consider only the highest spin assignments, namely ($13/2^-$) or ($15/2^-$.)  A similar high-energy transition at 2256~keV is observed in coincidence with the 429.1~keV $\gamma$-ray line and, thus, has to de-excite a newly observed level at 5462~keV which is likely to be ($15/2^-$,$17/2^-$).\\

\section{\label{sec:85As}$^{85}$As}

\begin{table}
\begin{center}
\begin{tabular}{c|c|c|c|c}
\hline
\hline
$E_{level},~keV $ & $J^\pi$& $E_{\gamma}$,~keV &  $I_{\gamma}$ & Coincidence \\
\hline
\hline
500.4 & ($7/2^-$)&500.4(2)& 28(3) & -\\
\hline
620.0 & ($7/2^-, 9/2^-$)&620.0(1)& 100(3) &793.6\\
\hline
796.6 & ($9/2^-$)&796.6(3)  & 138(19) &985.0\\
\hline
1413.6 & ($11/2^-$)&793.6(5) & 56(17) &620.0,  \\
 & & &  &711.0, 752.7\\
\hline
1781.6 & ($13/2^-$) &985.0(1) & 45(3) &796.6 \\
\hline
2113.2 & ($9/2^+, 13/2^-$)&1316.6(3) & 11(2) &796.6\\
\hline
\hline
 & &233.1(5) & 13(2) & - \\
\hline
 & &267.4(1) & 31(3) & - \\
\hline
 & &493.2(4) & 12(2) & - \\
\hline
 & &609.5(4) & 22(2) & - \\
\hline
 & &632.7(2) & 28(2) & - \\
\hline
 & &711.0(2) & 27(3) & 793.6 \\
\hline
 & & 752.7(3) & 12(2) & - \\
\hline
 & &788.8(4) & 25(3) & - \\
 \hline
 & &876.7(4) & 27(2) & - \\
\hline
 & &951.4(3) & 14(2) & - \\
\hline
 & &957.4(5) & 6(2) & - \\
\hline
 & &1219.4(4) & 12(2) & - \\
\hline
\hline
\end{tabular}
\caption{ Relative $\gamma$-ray intensities observed in coincidence with the isotopically-identified $^{85}$As, normalised to the 620.0 keV transition, along with the tentative spin assignments and observed coincident transitions. Despite its high intensity, the 796.6 keV transition has not been used to normalise the $\gamma$-ray intensities, since this transition is a triplet (see Figure~\ref{85As_gates}).}
\label{85As_tab}
\end{center}
\end{table}

Currently there is little information about the level structure of $^{85}$As. $\beta$-decay studies of $^{85}$Ge \cite{Korgul17} allowed to establish the first level scheme of $^{85}$As; however, very few spin assignments could be made. In our work we extend the proposed level scheme by introducing 5 new excited states.\par

The tracked Doppler-corrected $\gamma$-ray spectrum coincident with the $^{85}$As ion detection in VAMOS++ spectrometer is shown in figure \ref{85As_singles}. In total, 4$\cdot 10^5$ $^{85}$As events were identified. The full list of the $\gamma$-ray transitions observed in coincidence with $^{85}$As ions along with their intensities is presented in Table \ref{85As_tab}. \par

\begin{figure}
\centering
\includegraphics[width=0.99\linewidth]{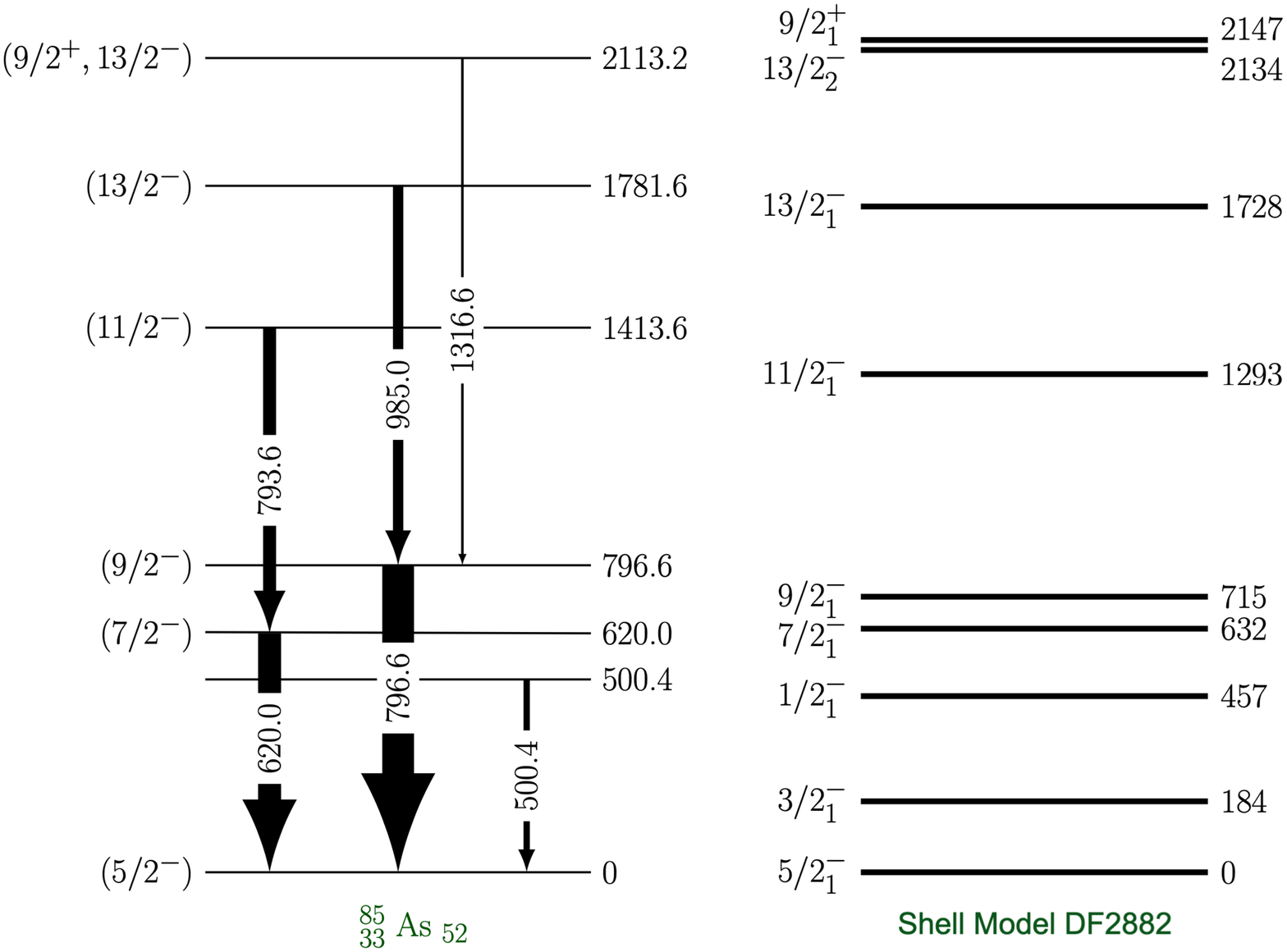} 
\caption{Left: partial level scheme of $^{85}$As, established in this work. Right: level scheme predicted by the LSSM calculations.}
\label{85As_levels_Nov2021}
\end{figure}

For the ground state of the $^{85}$As, (5/2$^-$) spin and parity was suggested in the $\beta$-decay study by Korgul et al. \cite{Korgul17} based on the systematics of $N$=52 isotopes. The 102~keV and 116~keV transitions were previously reported in Ref. \cite{Korgul17}. Due to the large Compton background originating both from $^{85}$As and its fission partner nuclei, the intensities of these transitions cannot be extracted from the present experimental data; however, it can be remarked that the excited states at 102~keV and 218~keV tentatively assigned in Ref. \cite{Korgul17} to be (3/2$^{-}$) and (1/2$^{-}$) respectively are, coherently with the yrast character of the reaction we used, only weakly populated in our experiment. The 500~keV transition observed in \cite{Korgul17} is also seen in the present work. No $\gamma$-rays are observed in coincidence despite its $\sim$20\% intensity, suggesting the direct population of the 500.4~keV level in fission. The 267.4~keV transition is also observed in the present work. However, since it does not appear in coincidence with the 233.1~keV transition described in Ref. \cite{Korgul17}, this must be a different $\gamma$-ray line.
The rest of the transitions are reported for the first time. The level scheme containing only the transitions that could be placed and were observed in the present work is presented in Figure~\ref{85As_levels_Nov2021} (left). \par

\begin{figure*}
\centering
\includegraphics[width=0.95\linewidth]{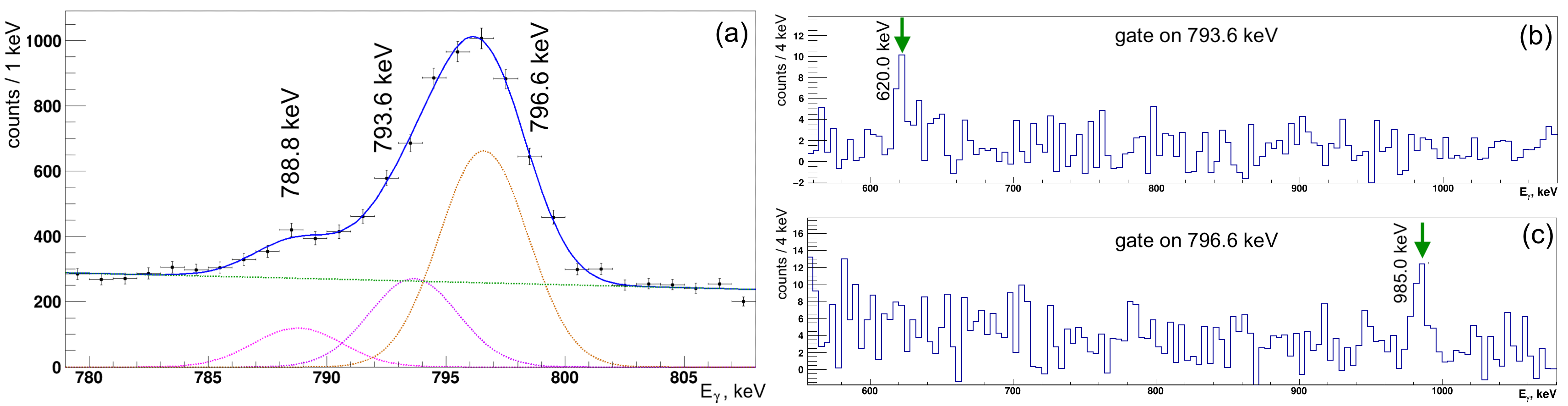} 
\caption{(a) Tracked $\gamma$-ray spectrum region seen in coincidence with $^{85}$As identified in VAMOS++, fitted with 3 Gaussian components (b) background-subtracted gate on the 793.6~keV transition (c) background-subtracted gate on 796.6~keV transition}
\label{85As_gates}
\end{figure*}

In the region of the spectrum near 795~keV, three $\gamma$-ray lines are seen, close in energy. To extract their intensities, the data were fitted with three Gaussian distributions and a linear background, with the mean values of 788.8(4), 793.6(5) and 796.6(3)~keV, corresponding to the $\gamma$-ray lines in coincidence with other transitions (see Figure~\ref{85As_gates}(a)).
The 793.6~keV transition is seen in coincidence with the 620.0~keV $\gamma$-ray line, while 796.6~keV $\gamma$-rays are in coincidence with the 985.0~keV line see Figure~\ref{85As_gates}(b,c).  Based on the relative intensities, the 620.0~keV and the 796.6~keV transitions are placed at the bottom of the level scheme. The 796.6 keV level is also fed from above by a 1316.6~keV transition. Both parallel $\gamma$-ray cascades are assumed to feed directly the ground-state.\par

Two more transitions 711.0 and 752.7~keV, are seen in coincidence with the 793.6~keV $\gamma$-ray line, but could not been placed in the level scheme due to low statistics.  The summary of all transitions corresponding to $^{85}$As seen in $\gamma$-$\gamma$ coincidences is given in Table~\ref{85As_tab}. \par

\section{\label{sec:87As}$^{87}$As}

\begin{figure*}
\centering
\includegraphics[width=0.95\linewidth]{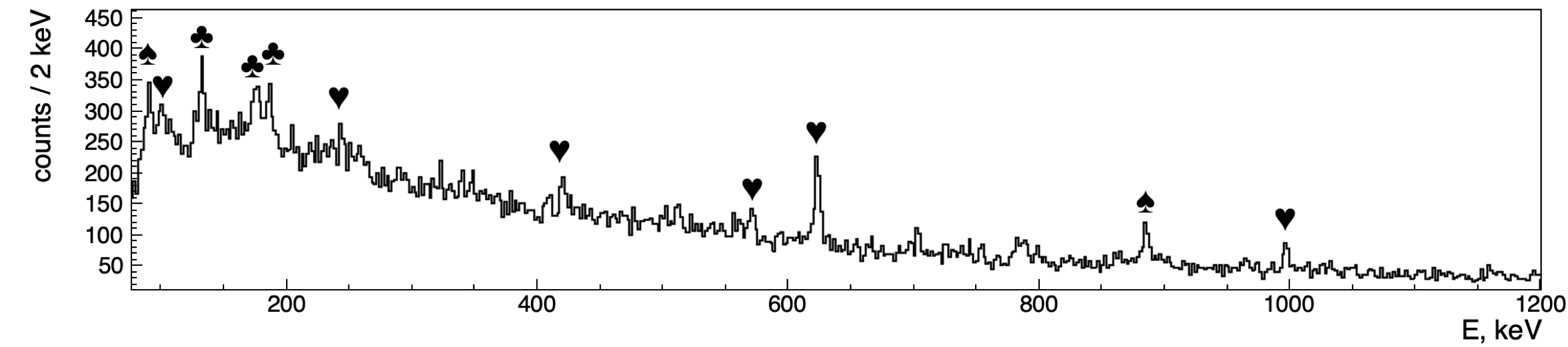} 
\caption{Tracked $\gamma$-ray spectrum in coincidence with $^{87}$As ions identified in VAMOS++.  ($\varheart$): $^{87}$As transitions reported for the first time in this work; ($\spadesuit$):  $^{87}$Se contaminant transitions; ($\clubsuit$) $^{86}$As contaminant transitions.}
\label{87As_singles}
\end{figure*}

\begin{figure}
\centering
\includegraphics[width=0.99\linewidth]{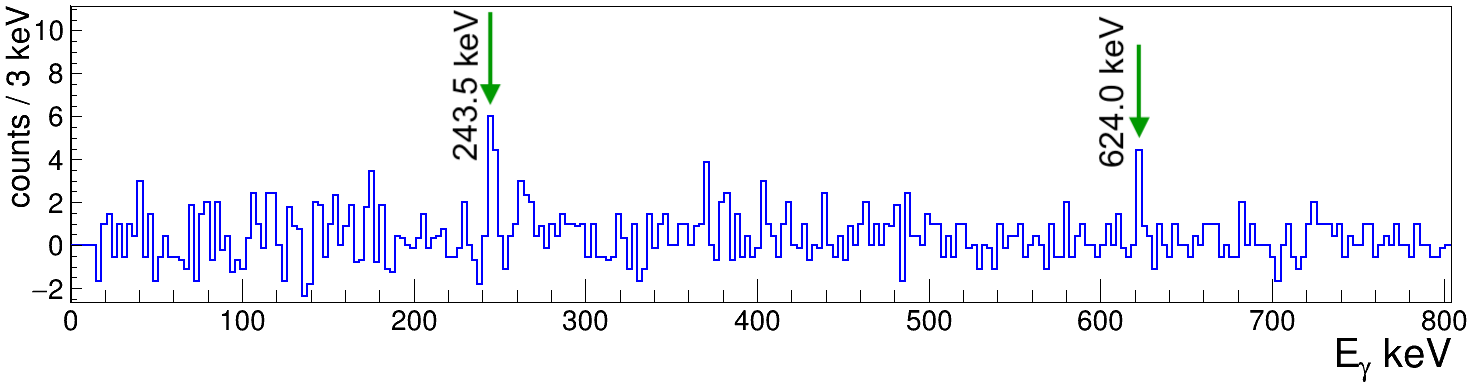} 
\caption{Tracked $\gamma$-ray spectrum observed in coincidence with the isotopically-identified $^{87}$As and the 997.3~keV $\gamma$-ray.}
\label{87As_gate997}
\end{figure}

Though the statistics available for $^{87}$As is very limited, with only 5$\cdot 10^4$ events identified in the VAMOS++ spectrometer, we could extract some spectroscopic data and propose the first level scheme for this nucleus. The ordering of the transitions is based on the coincidences and on the relative intensities of the observed $\gamma$-ray lines. Transitions at 101, 243.8, 420.7, 571.8, 624.0 and 997.3~keV have been attributed to $^{87}$As (see Table~\ref{87As_tab} and Figure~\ref{87As_singles}). Other peaks are attributed to the contaminants: $^{86}$As (133, 175, 187~keV) and $^{87}$Se (92, 886 and 1159~keV) which were both better produced in the present experiment. The 243.5~keV peak has a contribution from $^{86}$As. However, since it comes in coincidence with the 997.3~keV transition, this line is also attributed to $^{87}$As.

\begin{figure}
\centering
\includegraphics[width=0.99\linewidth]{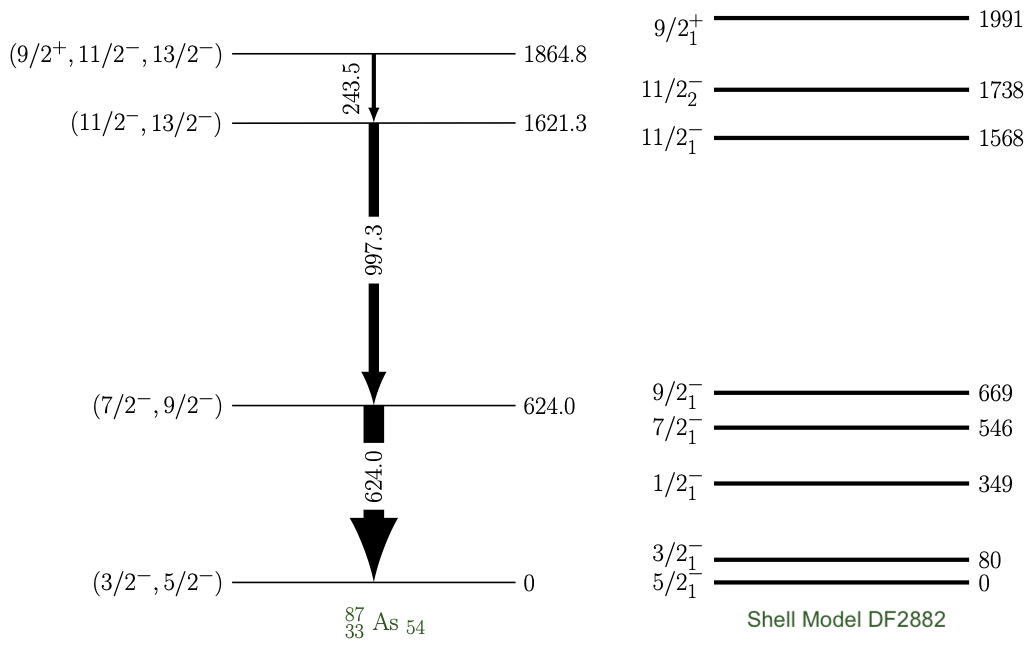} 
\caption{Left: level scheme of $^{87}$As, established in this work. Right: predictions from the LSSM calculations.}
\label{87As_levels_Nov2021}
\end{figure}

The proposed partial level scheme, based on the $\gamma$-$\gamma$ coincidences illustrated in Figure~\ref{87As_gate997} is presented in Figure~\ref{87As_levels_Nov2021} (left). The  101, 420.7 and 571.8~keV transitions were not placed in the level scheme since they were not seen in coincidence with any other transition. Figure \ref{87As_gate997} shows that the 243.5 keV and the 624.0 keV transitions are in coincidence with the 997.3 keV $\gamma$-ray line. Based on their relative intensities (see Table \ref{87As_tab}),  the cascade is ordered as shown in the level scheme. The spin assignments are tentative: the ground-state spin is assumed from the systematics to be ($3/2^-, 5/2^-$); the spins of the excited states are assumed to be yrast.  Therefore, the states at 624.0 keV and 1621.3 keV may be assigned ($7/2^-, 9/2^-$) and ($11/2^-$, $13/2^-$) spins and parities, respectively. The case of the highest lying state will be discussed in the following section.\\

\begin{table}[h]
\begin{center}
\begin{tabular}{c|c|c|c|c}
\hline
\hline
$E_{level},~keV $ & $J^\pi$ & $E_{\gamma}$,~keV & $I_{\gamma}$ & Coincidence \\
\hline
\hline
624.0 & ($7/2^-,9/2^-$)& 624.0(3) & 100(16) & 997.3   \\
\hline
1621.3 & ($11/2^-, 13/2^-$) & 997.3(4) & 50(9) & 243.5, 624.0   \\
\hline
1865 & ($9/2^+, 11/2^-,13/2^-$)& 243.5(6) &  - & 997.3   \\
\hline
\hline
 & & 101 &  - & -   \\
\hline
 & & 420.7(4) & 32(8) & -   \\
\hline
 & & 571.8(5) & 32(12) & -   \\
\hline
\hline
\end{tabular}
\caption{Relative $\gamma$-ray intensities observed in coincidence with the isotopically-identified $^{87}$As, normalized to the 624.0~keV transition, along with the tentative spin assignments and observed coincident transitions.}
\label{87As_tab}
\end{center}
\end{table}

\section{\label{sec:Discussion}Discussion}
In order to interpret the present experimental data, we have performed shell-model calculations
for the  $^{83-87}$As odd-mass isotopes. Following earlier works~\cite{Ni78-coreI,Ni78-coreII}, the shell-model valence space is spanned by the full $Z$=28-50 proton major shell and the full $N$=50-82 neutron major shell beyond $^{78}\text{Ni}$, namely, the  ($0f_{5/2}, 1p_{3/2}, 1p_{1/2}, 0g_{9/2}$) proton orbitals and the ($0g_{7/2}, 1d_{5/2}, 1d_{3/2}, 2s_{1/2}, 0h_{11/2}$) neutron orbitals. 
The set of Two-Body Matrix Elements (TBME), hereafter named DF2882 is  composed of the JUN45 interaction~\cite{JUN45} (with a slight $g_{9/2}$ single-particle energy adjustment) for the proton-proton interaction, the GCN5082 interaction~\cite{GCN5082-I,GCN5082-II} for the neutron-neutron interaction, and the matrix elements from Ref.~\cite{Ni78-coreI} for the proton-neutron interaction. For the neutron-neutron matrix elements originally defined for a $^{100}$Sn core, we apply a pairing reduction to incorporate the core polarization mechanism differences between $^{78}$Ni and $^{100}$Sn, as detailed in reference~\cite{Core-pol}. This shows to be a major improvement in the 2$^+$ energies of the $N$=52 isotones over the previous work from Ref.~\cite{Ni78-coreI}.
Lastly, the cross-shell monopole matrix elements have been constrained to reproduce the spectroscopy evolution with neutron filling up to $N$ = 56. In particular, constraints were put on the evolution of the ${9/2}^+$ and ${13/2}^+$ states in rubidium and yttrium isotopes, as well as for the evolution of the first excited 0$^+$ state from $^{90}$Zr to $^{96}$Zr.
The overall agreement obtained for the spectroscopy of neutron-rich nuclei beyond $N$ = 50 is very good.

\subsection{$^{83}$As}
The chosen valence space only allows proton excitations for the semi-magic isotope $^{83}$As. Negative parity states involve predominantly the $f_{5/2}$ and $p_{3/2}$ orbitals. For the positive-parity states, a proton must be promoted to the $g_{9/2}$ orbital. The theoretical level scheme shown in Figure \ref{83As_levels_theor} (right) reproduces fairly well the experimental one up to the excitation energy $E^*\sim$2 MeV. In particular, the 1193.7 keV and the 1525.7 keV levels may be assigned spin-parity ($1/2^-$) and ($7/2^-$) based on the similarities with the 1191 keV and 1427 keV levels in $^{85}$Br \cite{Nyako1}, as well as the LSSM predictions (1317 keV and 1425 keV), respectively. At a higher excitation energy, the competition with particle-hole (p-h) excitations across the $N$ = 50 shell gap appears. Such states have been identified recently in $^{81}$Ga \cite{Dud81Ga} at excitation energies starting around $\sim$ 2.5 MeV. From the $N$ = 49 systematics~\cite{NNDC}, the lowest positive states across the $N$ = 50 gap show a minimum around $^{81}$Ge and $^{83}$Se. Therefore, p-h excitations are probably expected in the energy range of those observed in $^{81}$Ga from $\sim$ 2.5  MeV.  Calculations within the $pf-sdg$ extended valence space predict the $13/2^-_1$-$15/2^-_1$ doublet located at 2.50 MeV and 2.95 MeV respectively, while higher-spin states lie higher in excitation energy.

The lowest-lying positive-parity state, 9/2$^+$, is predicted at 2379 keV, which could correspond to one of the states drawn in green in Figure \ref{83As_levels_theor} (left), lying around this energy and feeding the 9/2$^-$ state at 1542.6 keV.

\subsection{$^{85}$As}
For the vast majority of states, and for the dominant configurations, the two valence neutrons in  $^{85}$As are predicted by LSSM calculations to appear in the $d_{5/2}$ orbital.  The configurations are much more mixed than in $^{83}$As, as expected. The state at 500.4 keV decaying directly to the ground state with a moderate $\gamma$-ray intensity,  may correspond to (3/2$^-$), since the (1/2$^-$) state is less likely to be directly populated in the studied reaction. The two states at 620.0 keV and 796.6 keV drain the full decay flux. They are thus assigned (7/2$^-$) and (9/2$^-$), in good agreement with the LSSM predictions (632 keV and 715 keV, respectively). The states at 1413.6 keV and 1781.6 keV decay both with about 50\% of the intensity of the 620.0 keV transition and should correspond to the (11/2$^-$) and (13/2$^-$) states, respectively. The similarity between $^{85}$As and $^{87}$Br \cite{Nyako1} for the (7/2$^-$), (9/2$^-$) and (11/2$^-$) states is striking. Finally, the higher-lying state at 2113.2 keV decaying to the (9/2$^-$) state, could correspond to the second (13/2$_2^-$) state or the (9/2$^+$) state.

\subsection{$^{87}$As}
Similarly to $^{85}$As, the dominant configurations in $^{87}$As are predicted to include four neutrons in the $d_{5/2}$ orbital. The configuration mixing is even stronger in $^{87}$As. A correspondence is found between the yrast states predicted by the LSSM calculations and the experimentally observed ones at 624.0 keV and 1621.3 keV excitation energy. The state at 1864.8 keV could correspond to a negative-parity state (11/2$^-_2$) or (13/2$^-$), or to the (9/2$^+$) state. With the latter assignment, the hypothesis that the 9/2$^+$ state in $^{83}$As lies around 2600 keV and that the highest state observed in $^{85}$As is a 9/2$^+$ state,  the systematics of the  9/2$^+$ states in the $N$=50, 52, 54 neutron-rich isotopes for As ($Z$=33), Br ($Z$=35), Rb ($Z$=37) and Y ($Z$=39) shows a similar smooth trend well reproduced by LSSM calculations (see section \ref{ShellEvol} and corresponding Figure~\ref{ninehalf}).

\subsection{Shell evolution}
\label{ShellEvol}
\begin{figure}[H] 
  \centering
  \includegraphics[width=0.75\linewidth]{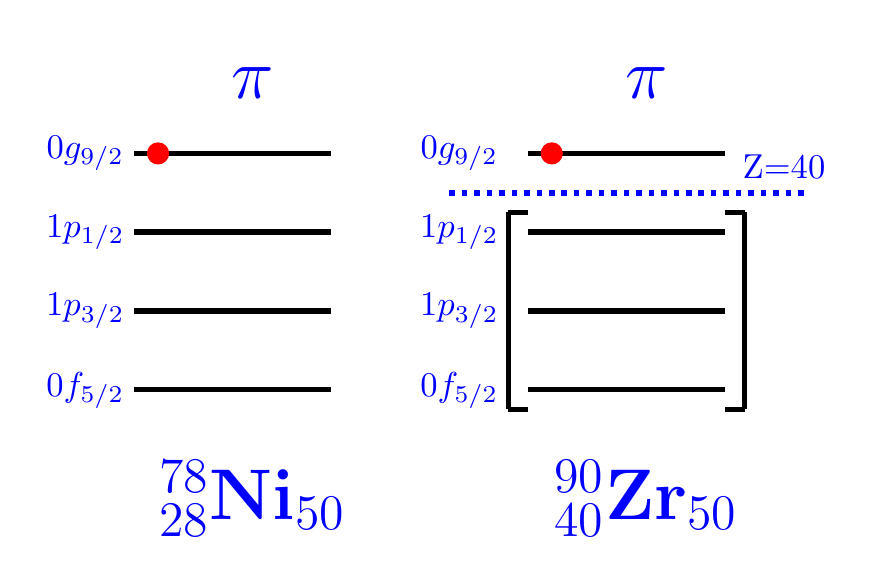}
   \includegraphics[width=0.75\linewidth]{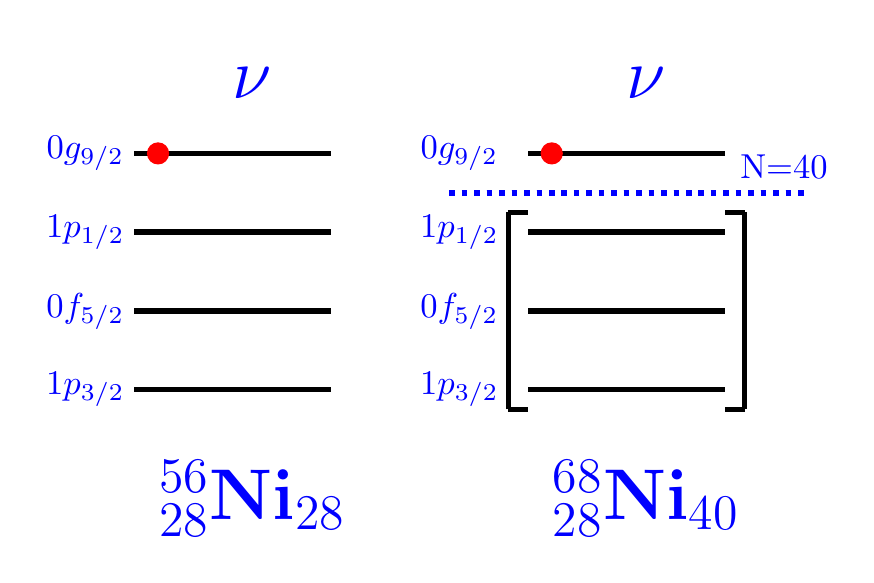} 
  \caption{Valence mirror symmetry in $N$=50 isotones (top) and $Z$=28  isotopes (bottom)}
  \label{mirror}
\end{figure}

\begin{figure*}
\centering
\includegraphics[width=0.75\linewidth]{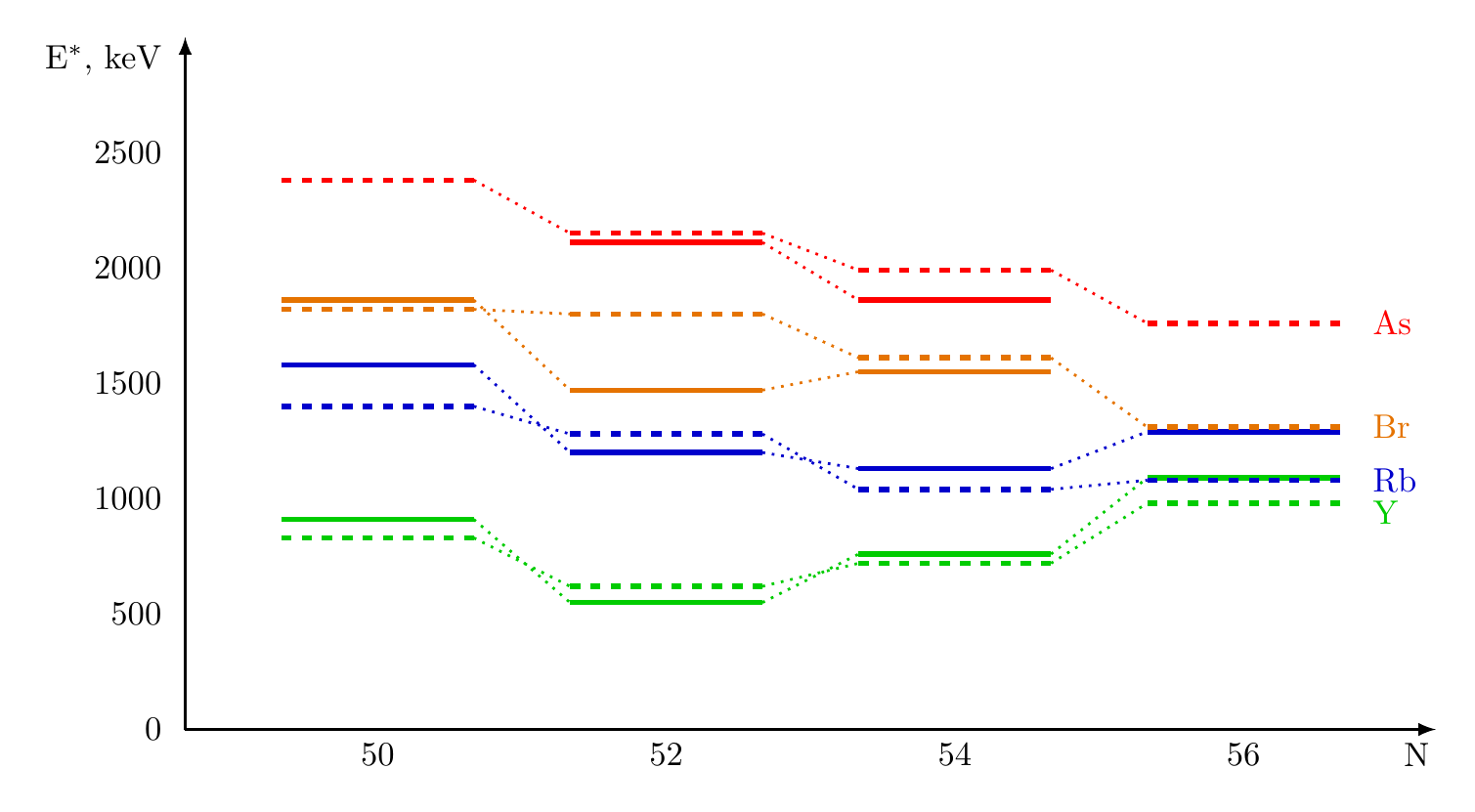} 
\caption{Excitation energy of the 9/2$^+$ states in As ($Z$=33), Br ($Z$=35), Rb ($Z$=37) and Y ($Z$=39) isotopes. The solid lines represent experimental values from~\cite{Nyako1, Nyako2, Rubidium1, Rubidium3} and the tentative assignments from this work. The dashed lines represent the predictions of the LSSM calculations.}
\label{ninehalf}
\end{figure*}

In addition to the observation of rotational motion developing for open-shell cases, the present data and their theoretical interpretation allow to sketch the evolution of the proton $g_{9/2}$ orbital along the semi-magic $N$=50 isotonic chain.
Figure~\ref{ninehalf} shows the identified 9/2$^+$ states in the region. Inspection of the corresponding wave functions shows that all these states are  built on a proton $g_{9/2}$ excitation. 
 $^{85}$As and $^{87}$As represent the most neutron-rich cases where their placement has been suggested in the vicinity of the $N$=50 isotonic chain. As expected, the overall trend shows a lowering the 9/2$^+$ state along $N$=50 isotonis chain with increasing valence proton holes  towards a mininum  at mid-shell between magic $^{78}$Ni and $^{90}$Zr.  
    The excellent  reproduction of the 9/2$^+$ systematic by the present effective interaction gives confidence in the extrapolated excitation energy value of the $g_{9/2}$ orbital location in $^{79}$Cu to 3.95 MeV. Such an estimate can also be inferred from the valence mirror symmetry between $Z$=28 isotopes and $N=50$ isotones as shown in Figure~\ref{mirror}. At first order, the experimental $Z$=40 shell gap $^{90}$Zr appear to be identical to the experimental $N$=40 shell gap in $^{68}$Ni $\sim$ 3.2 MeV, implying that the $g_{9/2}$ single-proton value in $^{79}$Cu should be identical to the $g_{9/2}$ single-neutron value in  $^{57}$Ni. This latter value  of 3.7 MeV  from~\cite{NNDC} appears to be fully compatible with the present extrapolation  
    and provides an additional benchmark for single-particle states in the vicinity of $^{78}$Ni.

\subsection{Rotational motion}
\begin{figure}[H] 
\hskip -20pt
  \includegraphics[width=1\linewidth]{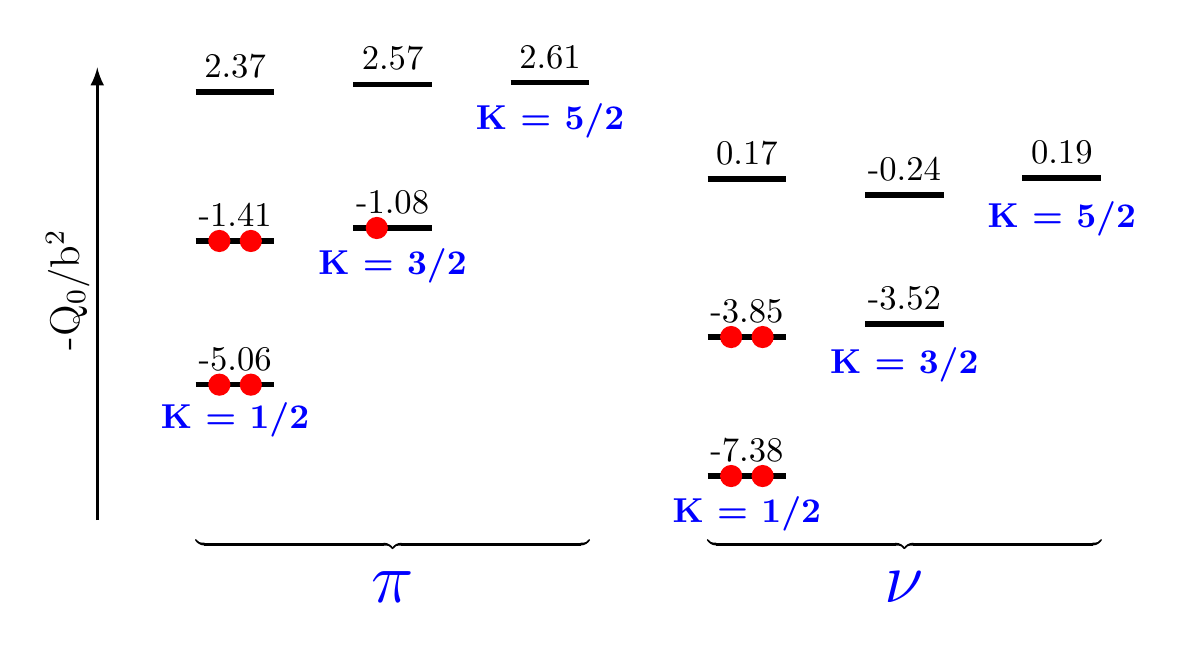}
  \caption{Zuker-Retamosa-Poves diagrams for the pseudo-SU3 proton orbitals (left) and neutron orbitals (right) limits in the case of $^{87}$As}
  \label{ZRP}
\end{figure}

\begin{figure}
  \centering
  \includegraphics[scale=0.32]{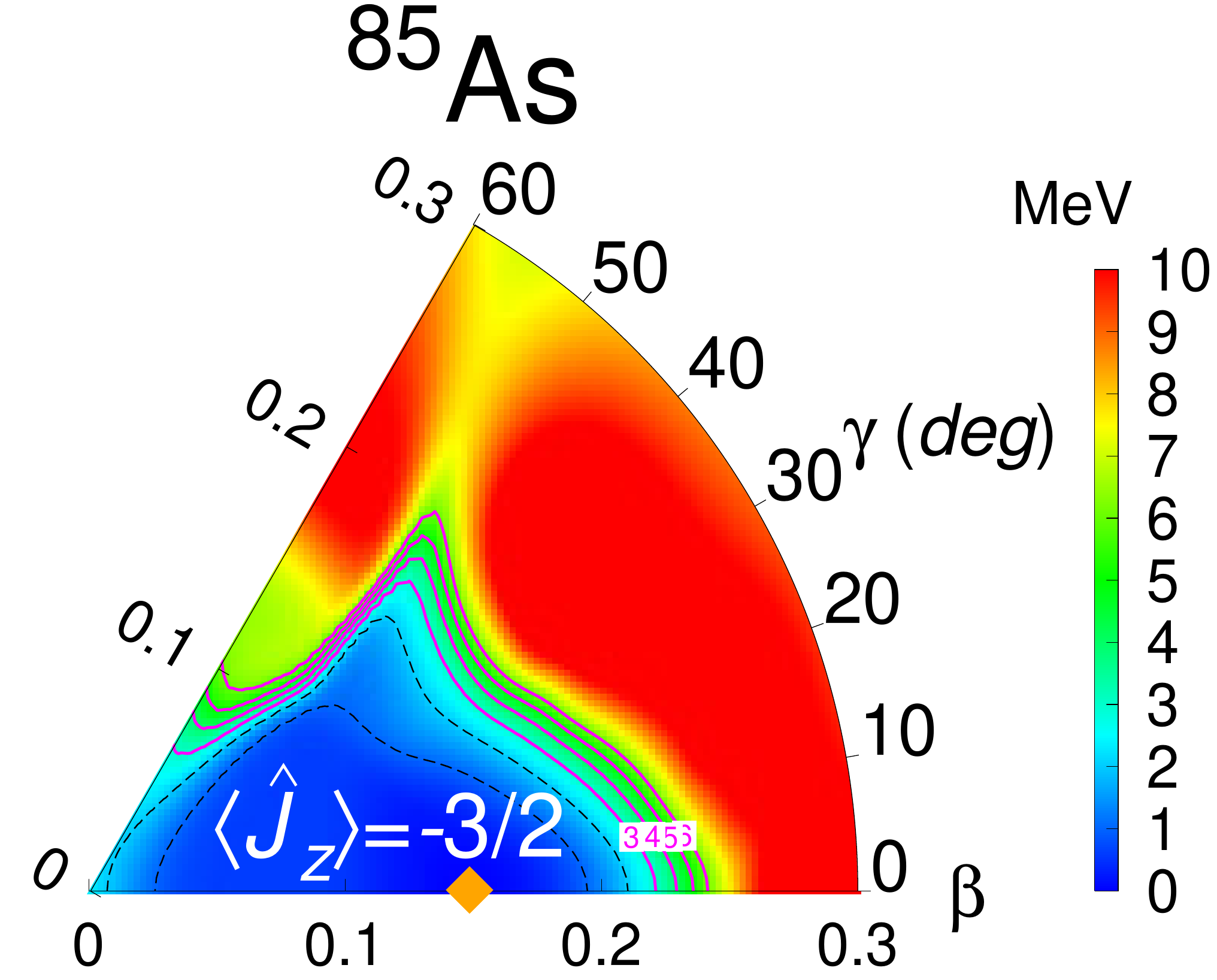}
  \caption{Potential energy surface (PES) calculated for $^{85}$As with the treatment of odd-mass nuclei by means of a cranking method following Ref.~\cite{Kasuya2021}. The cranked component corresponds to that of the HF minimum $\langle \hat J_z\rangle=-3/2$. The yellow diamond symbol represents the HF minimum.}
  \label{pesAs85}
\end{figure}

\begin{figure}[H]
  \centering
  \includegraphics[width=0.9\linewidth]{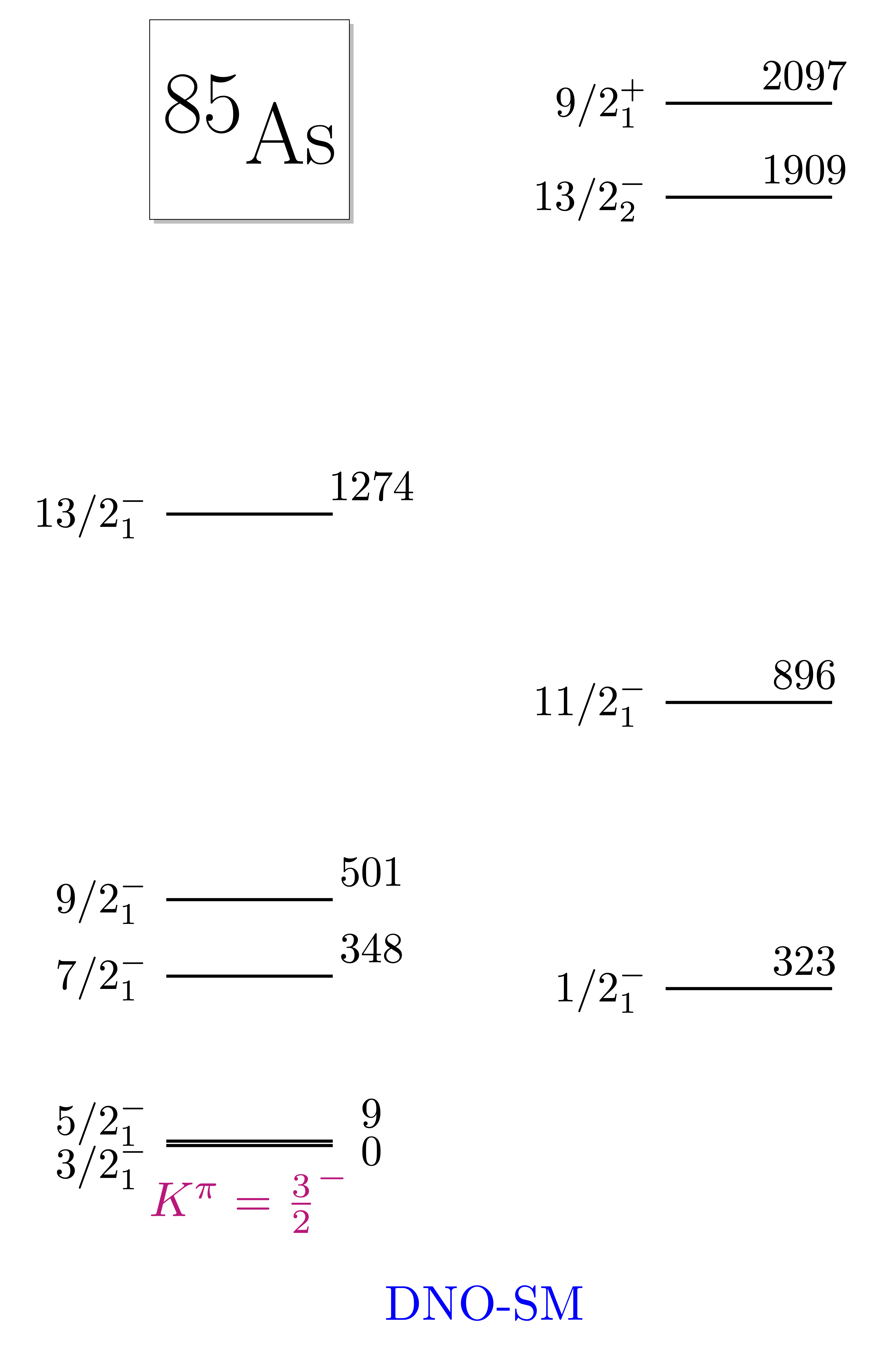}
  \caption{DNO-SM calculations for $^{85}$As, using 45 deformed HF states (see more details in Ref.~\cite{DNOSM}).}
  \label{specAs85}
\end{figure}

\begin{figure*}
  \includegraphics[scale=0.3]{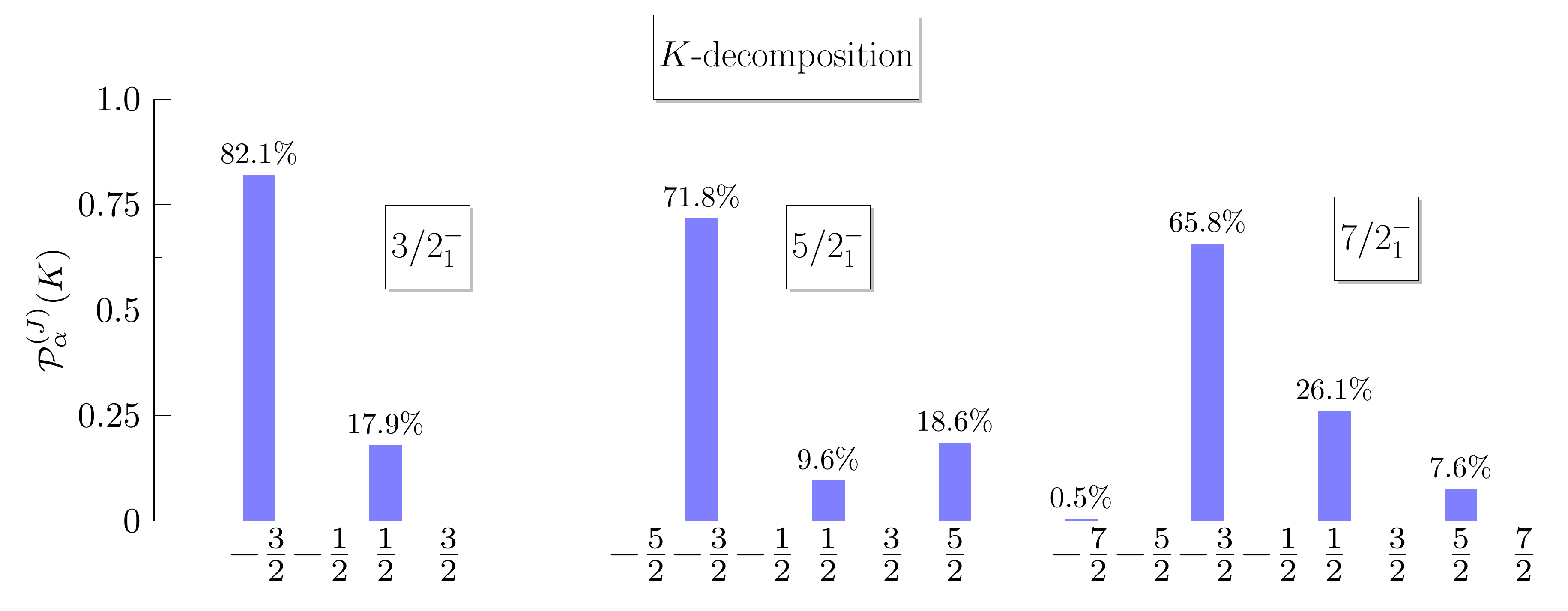}
  \includegraphics[scale=0.3]{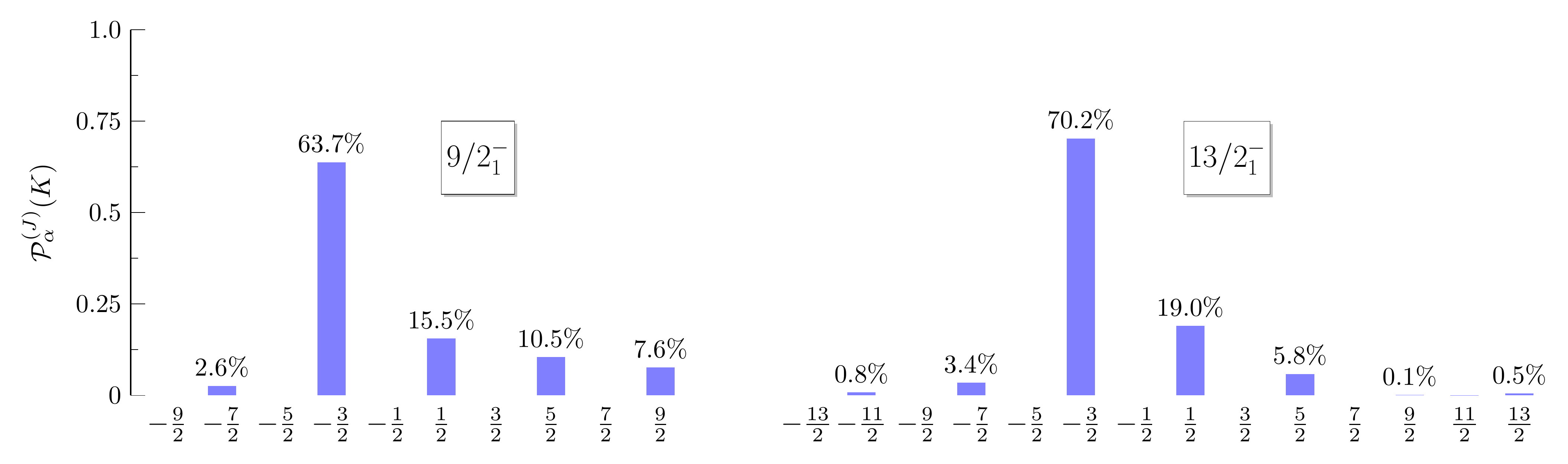}
  \includegraphics[scale=0.3]{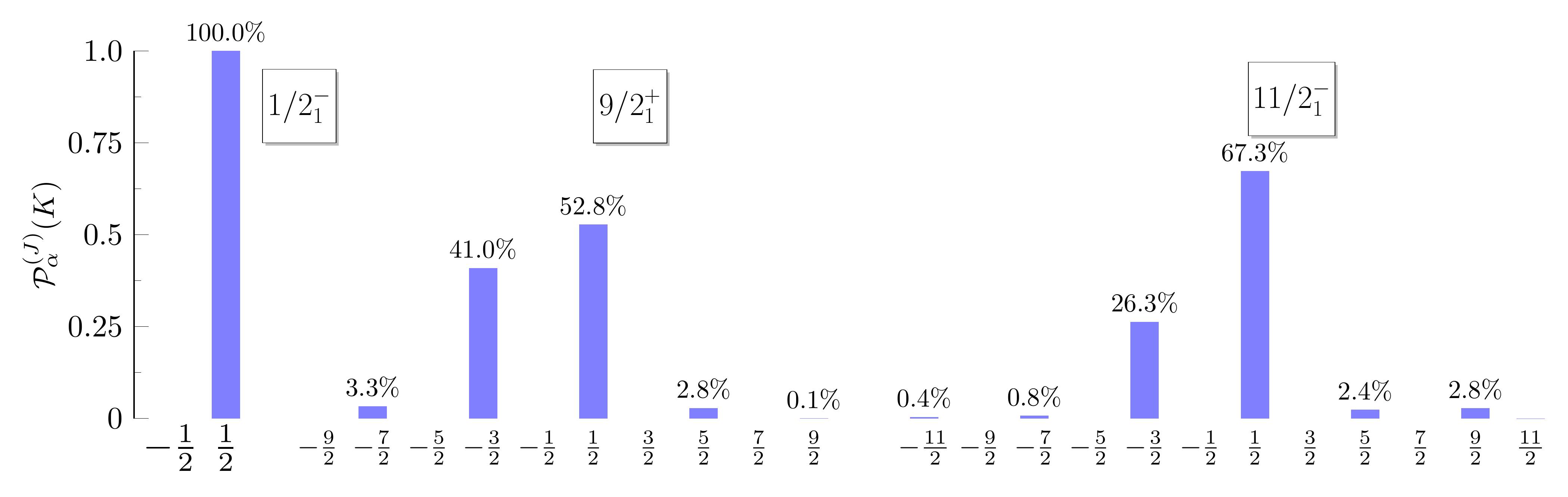}
  \includegraphics[scale=0.3]{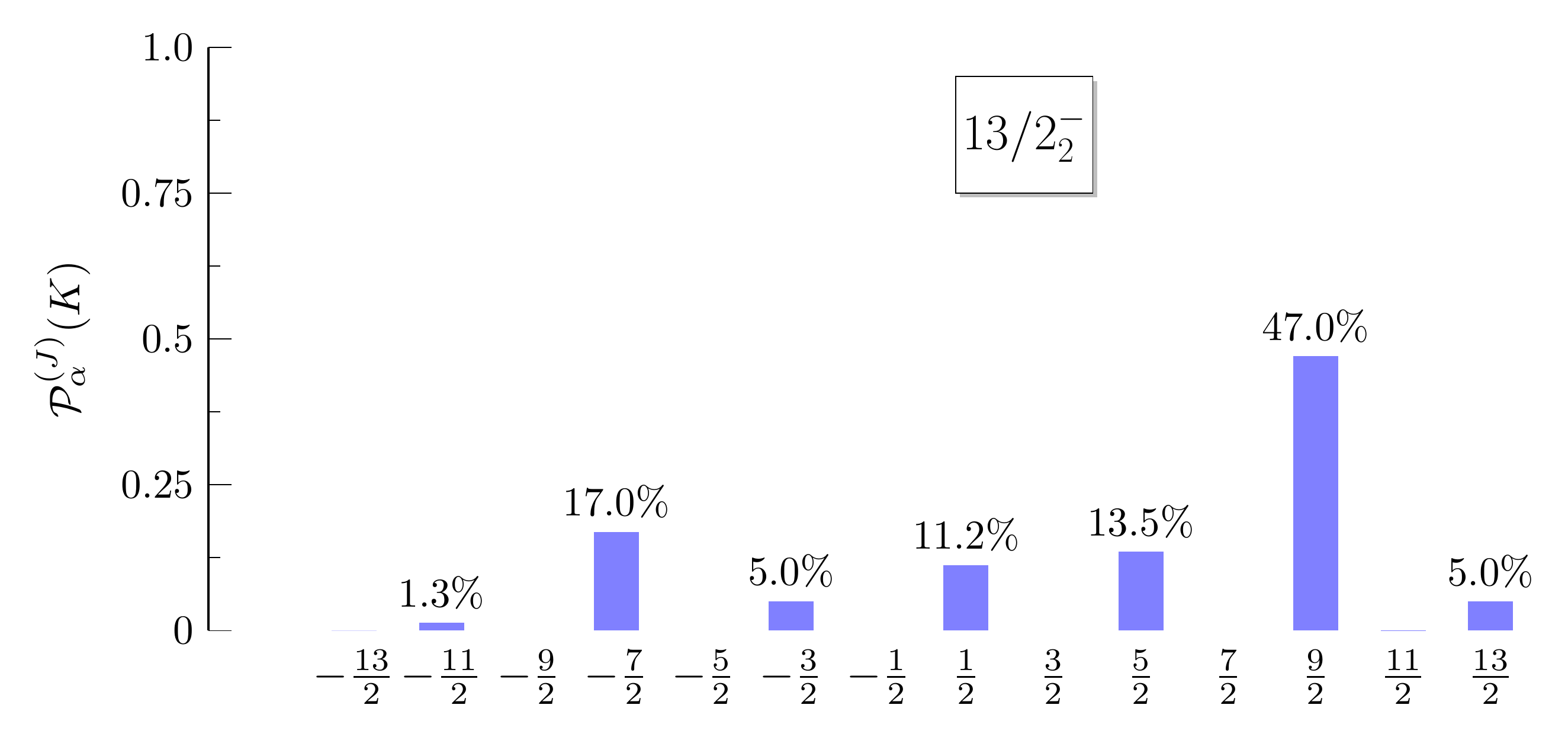}
  \caption{Wave functions content in $K$-quantum numbers of $^{85}$As. For the various states $\mathcal P_\alpha^{(J)}(K)$ (normalized to unity) is the contribution of a component $K$ in the given state $J_\alpha$.}
  \label{K_structure_As85}
\end{figure*}

\begin{figure}[H]
  \centering
  \includegraphics[scale=0.32]{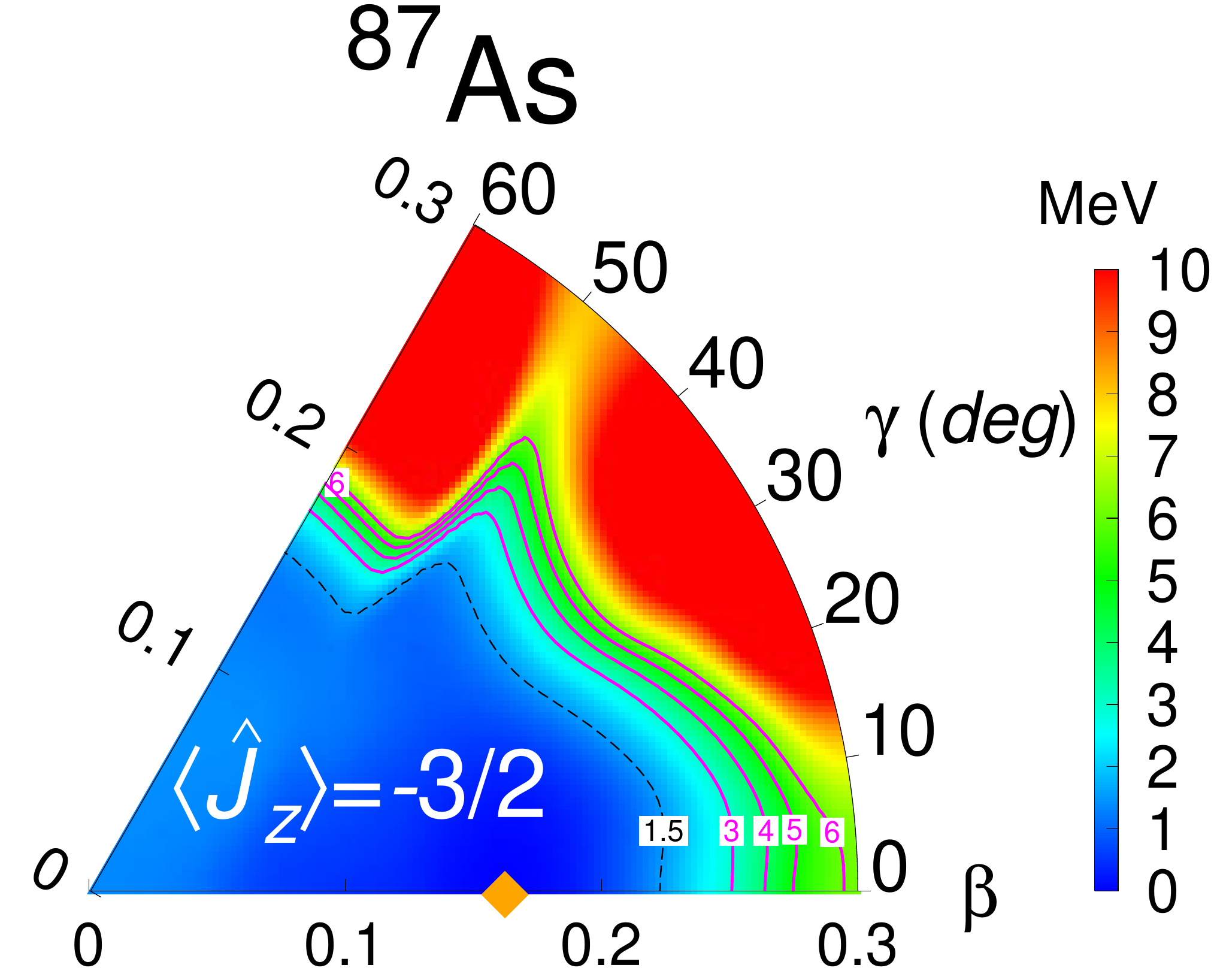}
  \caption{PES for $^{87}$As obtained with the same method as in Fig.~\ref{pesAs85}.}
  \label{pesAs87}
\end{figure}

\begin{figure}
  \centering
  \includegraphics[width=0.9\linewidth]{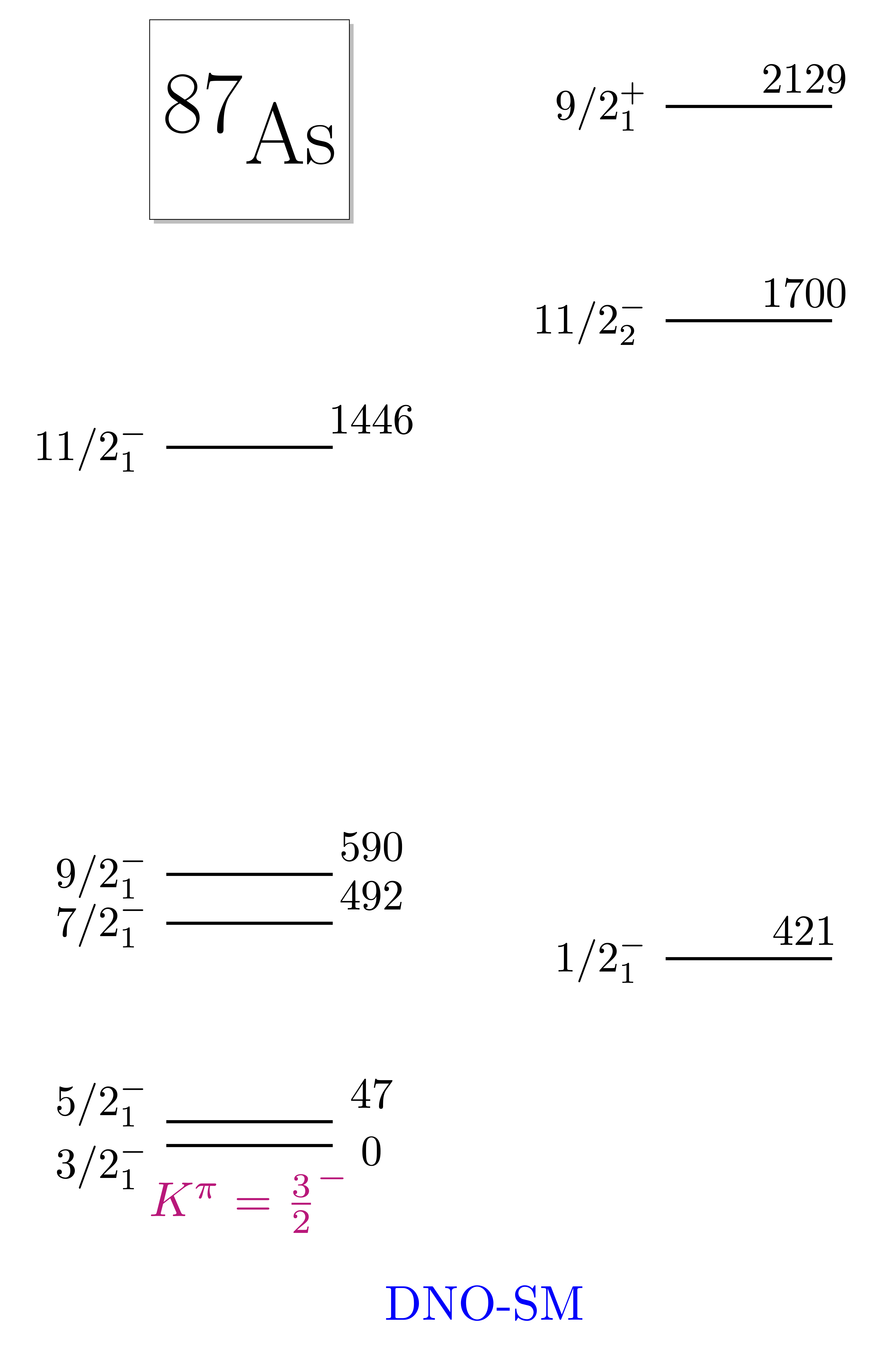}
  \caption{DNO-SM calculations for $^{87}$As, using 50 deformed HF states (see more details in Ref.~\cite{DNOSM}).}
  \label{specAs87}
\end{figure}

\begin{figure*}
  \includegraphics[scale=0.3]{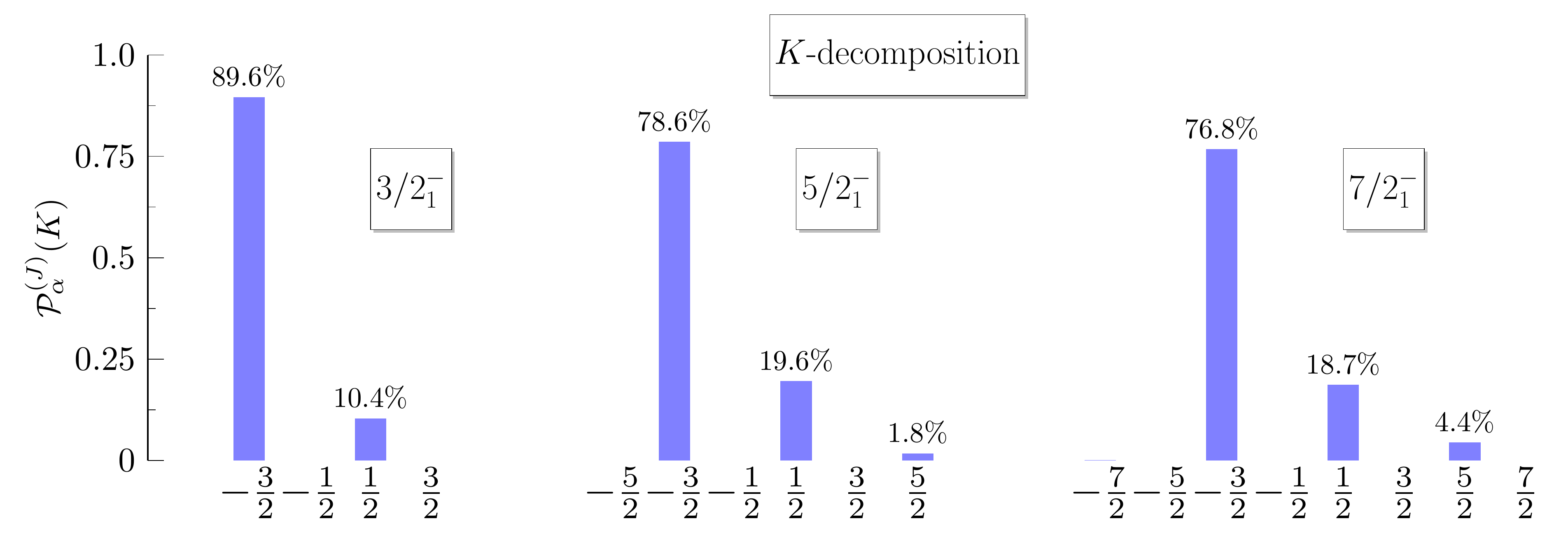}
  \includegraphics[scale=0.3]{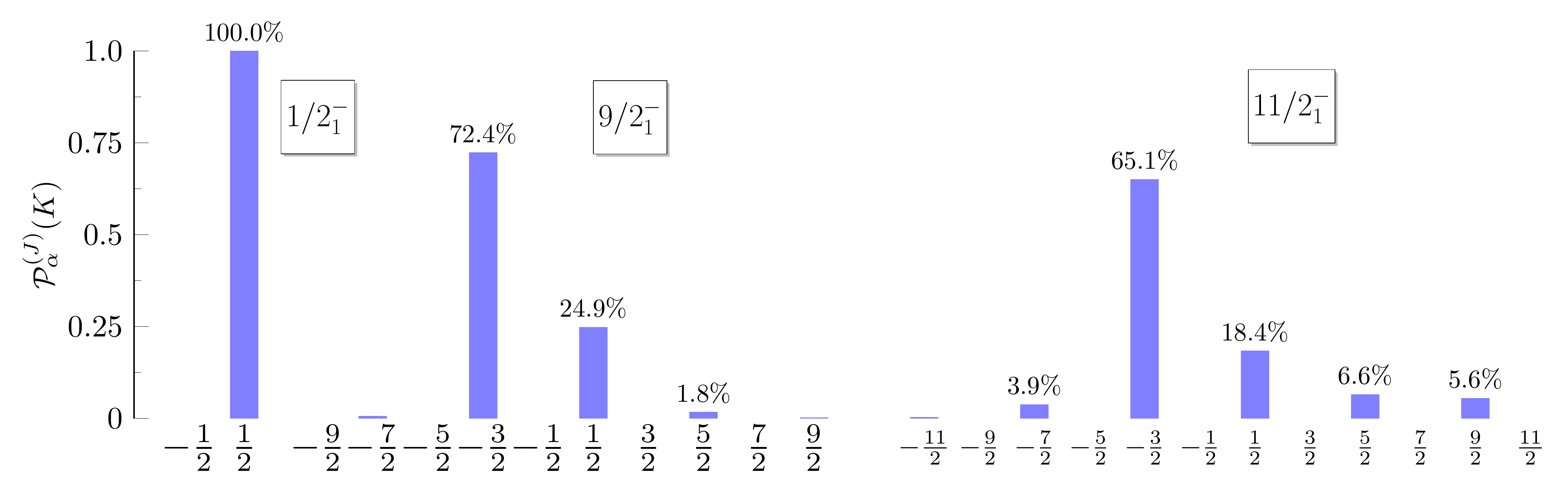}
  \includegraphics[scale=0.3]{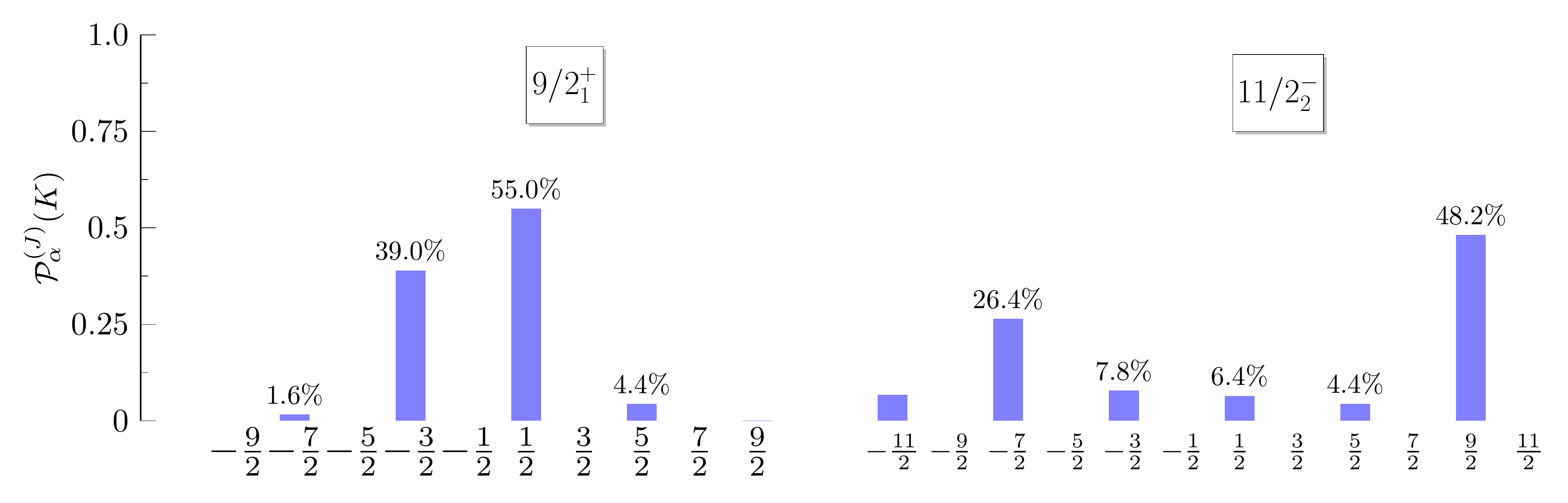}
  \caption{Same analysis in $K$-quantum numbers of $^{87}$As states as in Fig.~\ref{K_structure_As85}.}
  \label{K_structure_As87}
\end{figure*}

\begin{figure*}
  \includegraphics[scale=0.235]{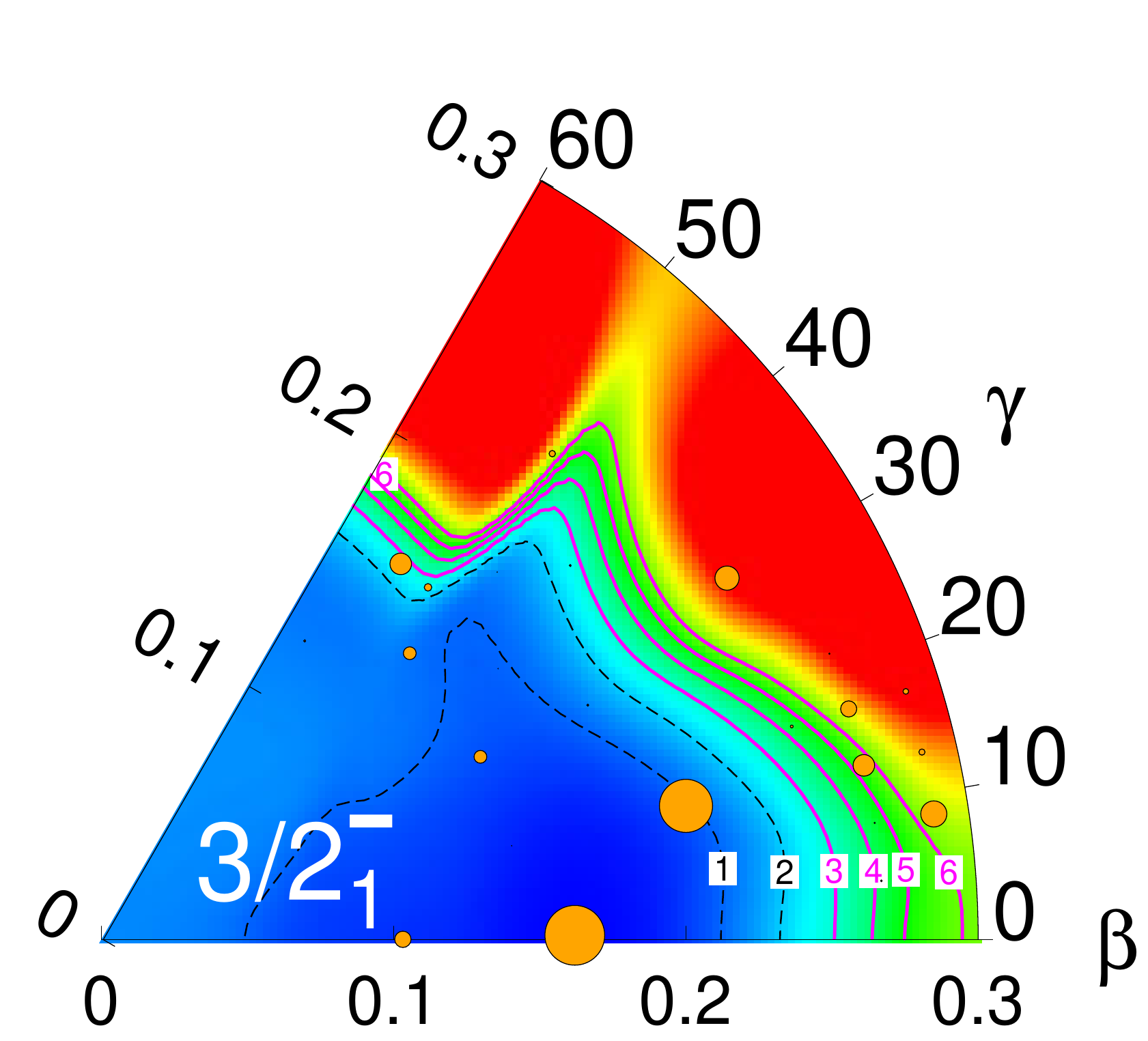}
  \includegraphics[scale=0.235]{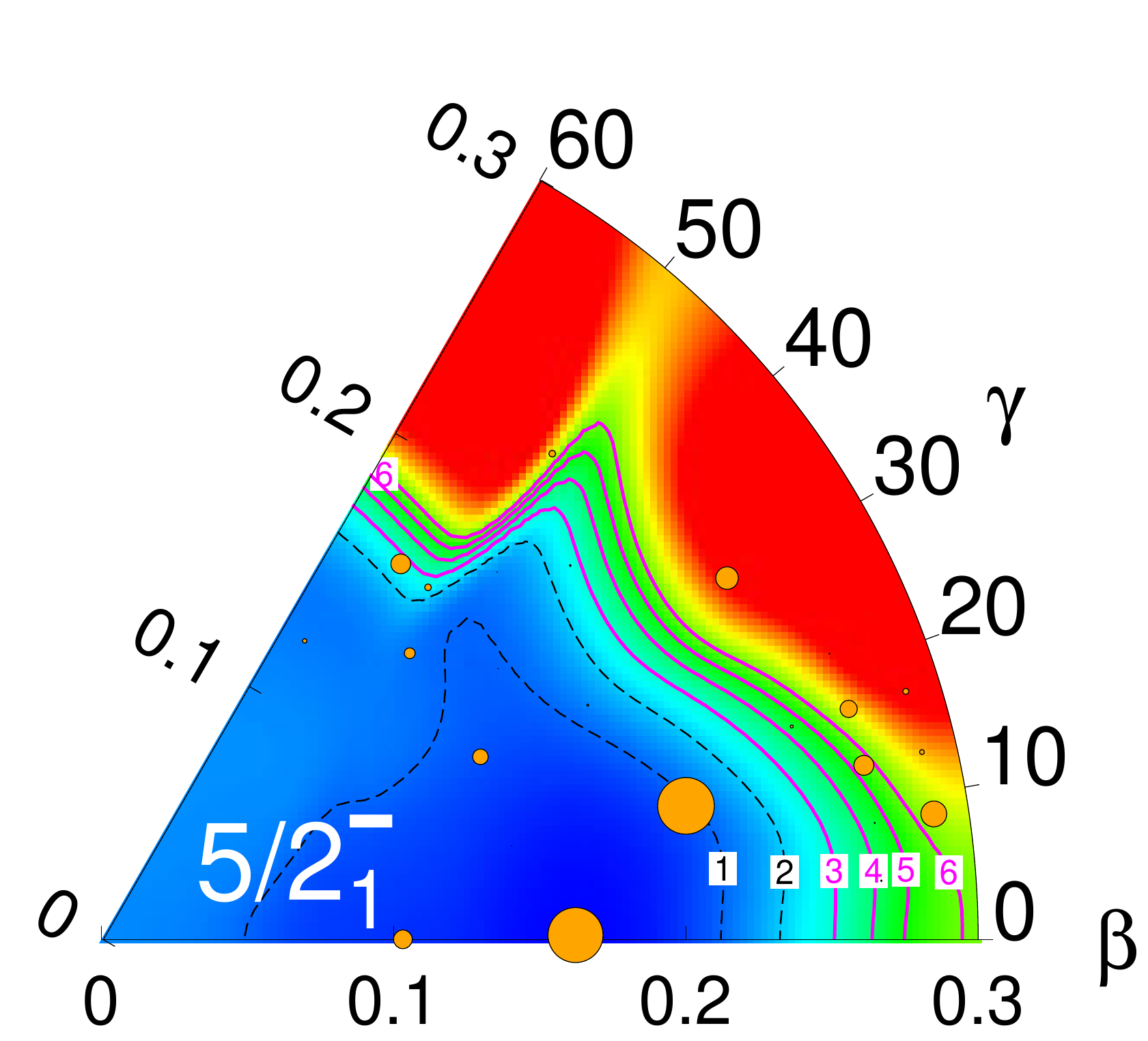}
  \includegraphics[scale=0.235]{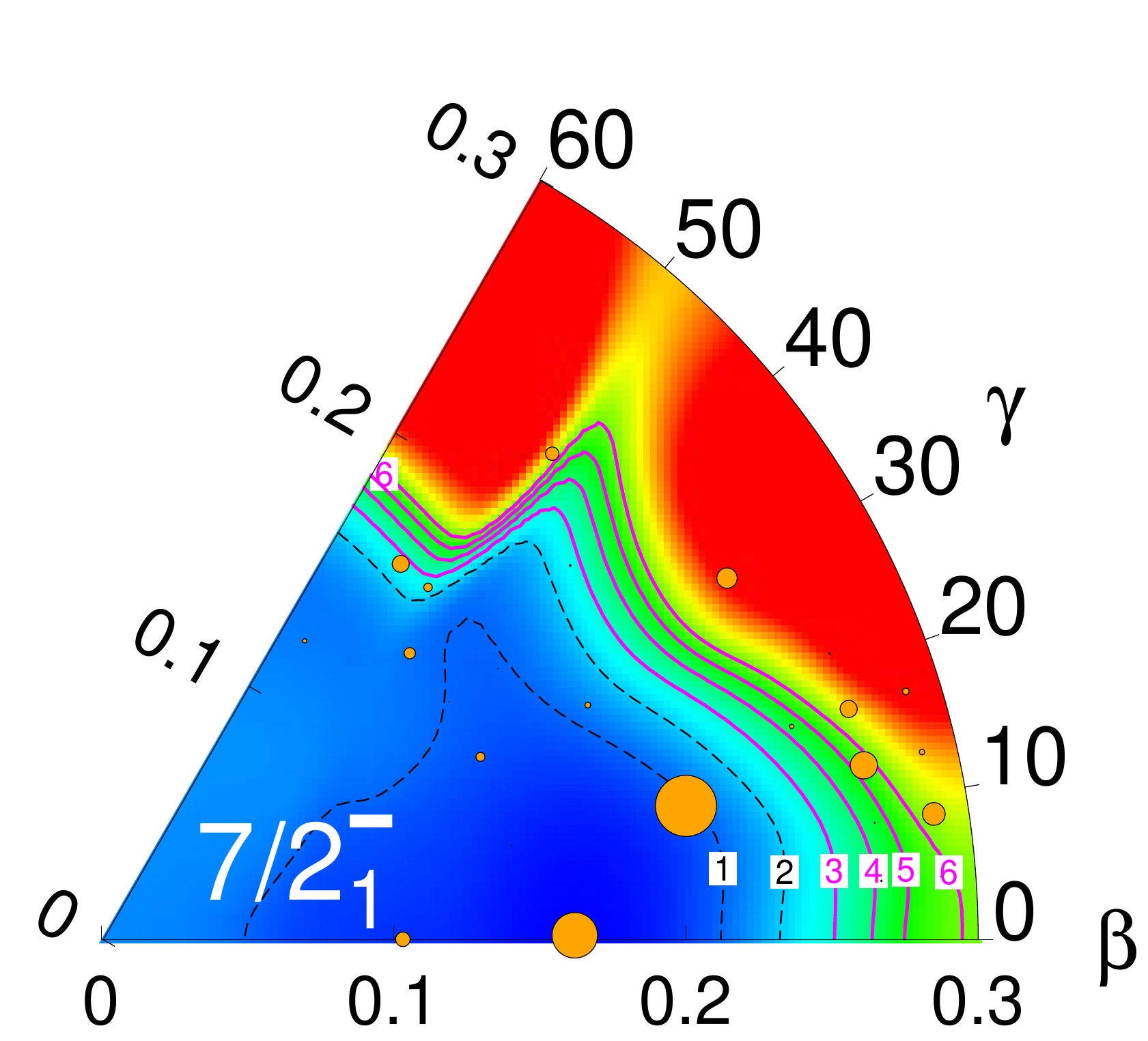}
  \centering 
  \includegraphics[scale=0.235]{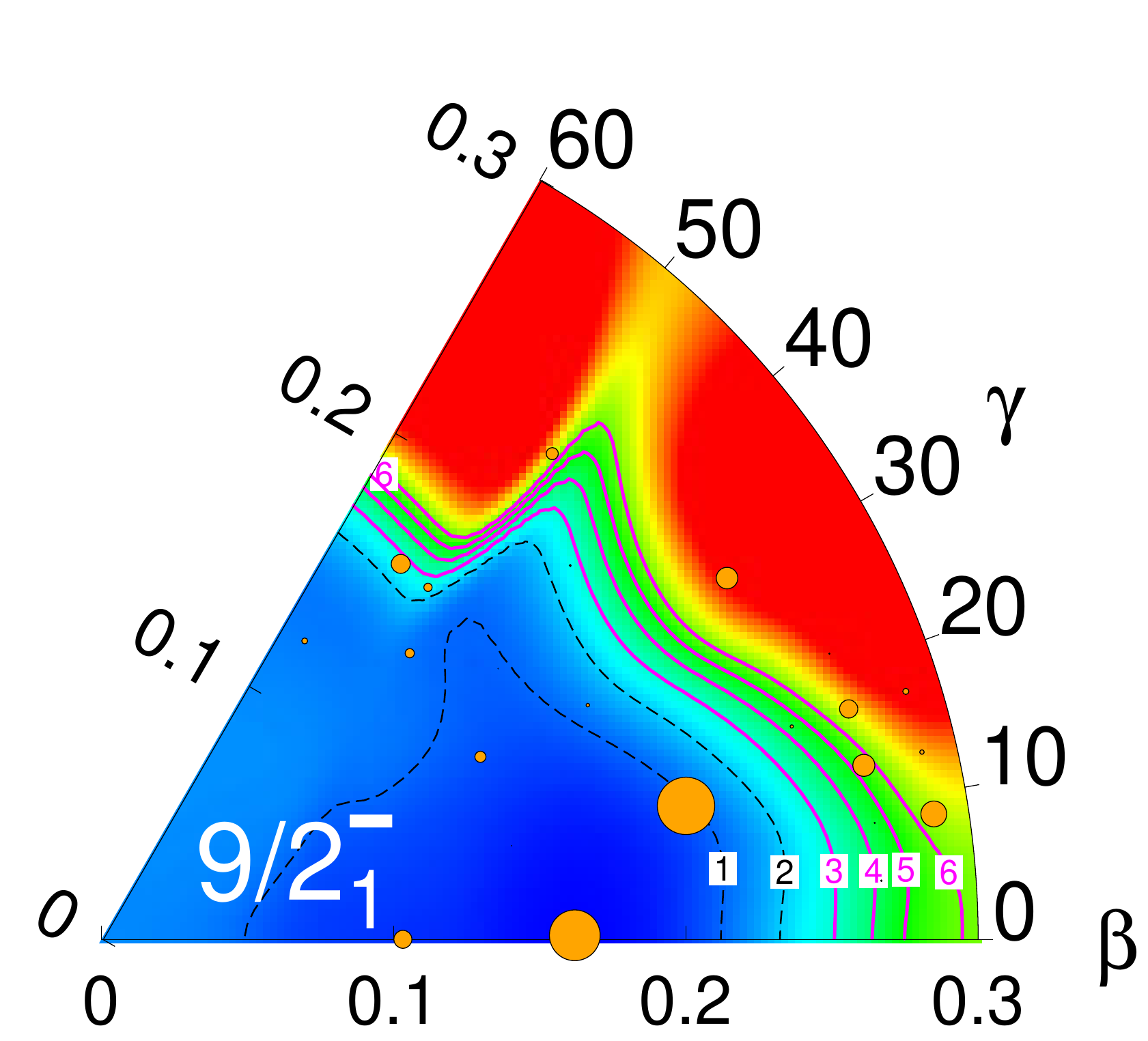}
  \includegraphics[scale=0.235]{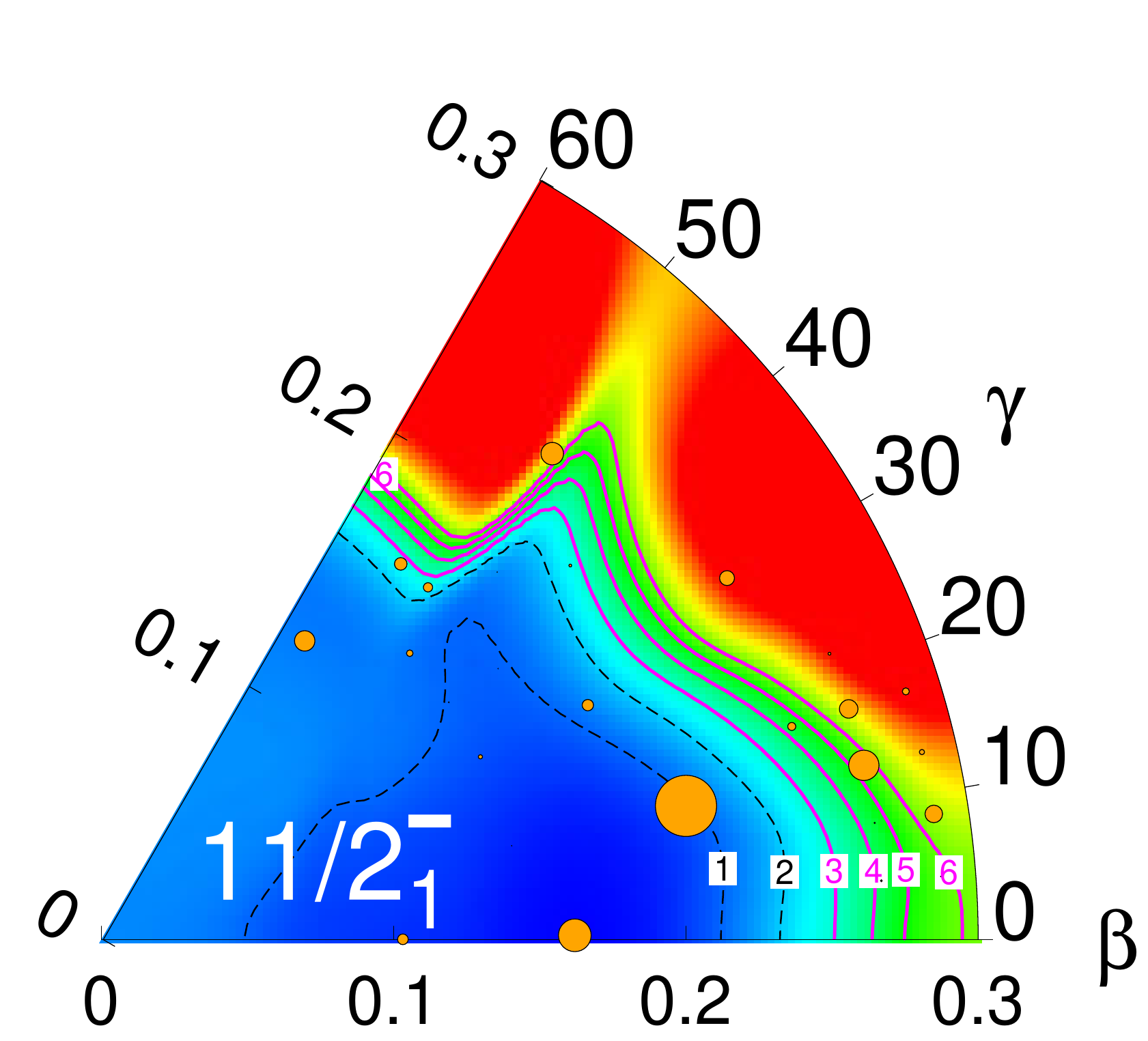}
  \hspace{5pt}
  \includegraphics[scale=0.235]{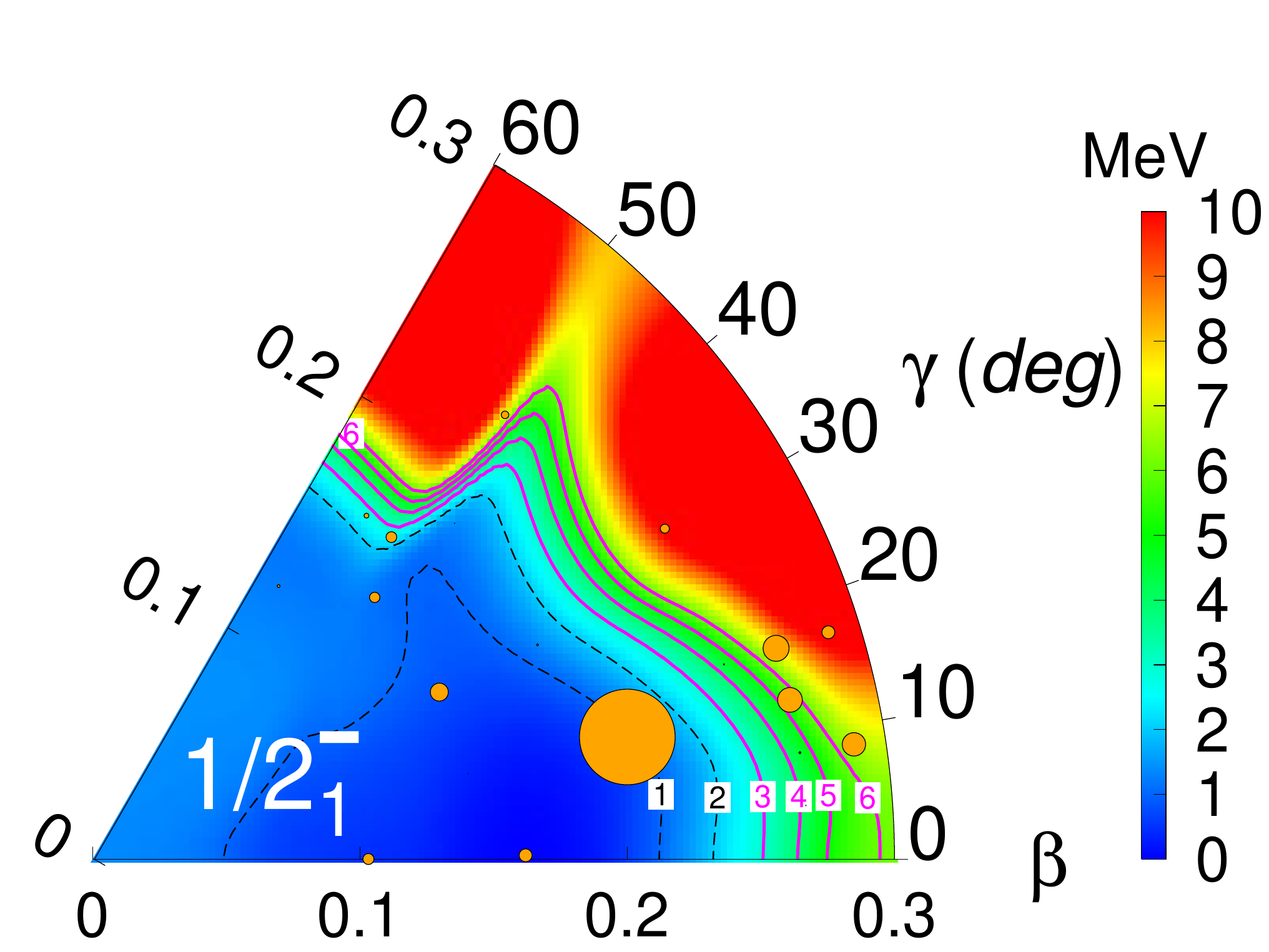}
  \includegraphics[scale=0.235]{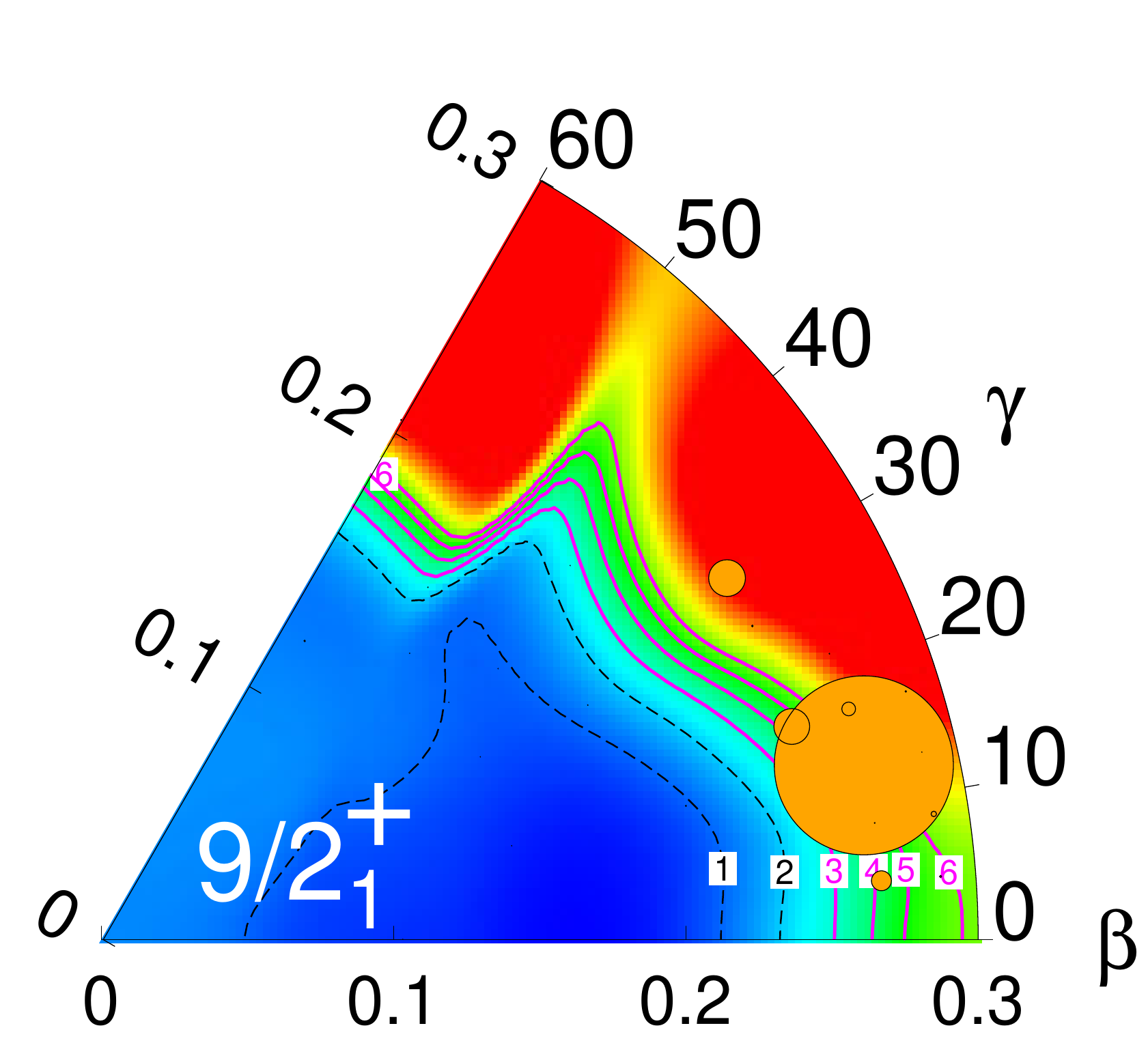}
  \includegraphics[scale=0.235]{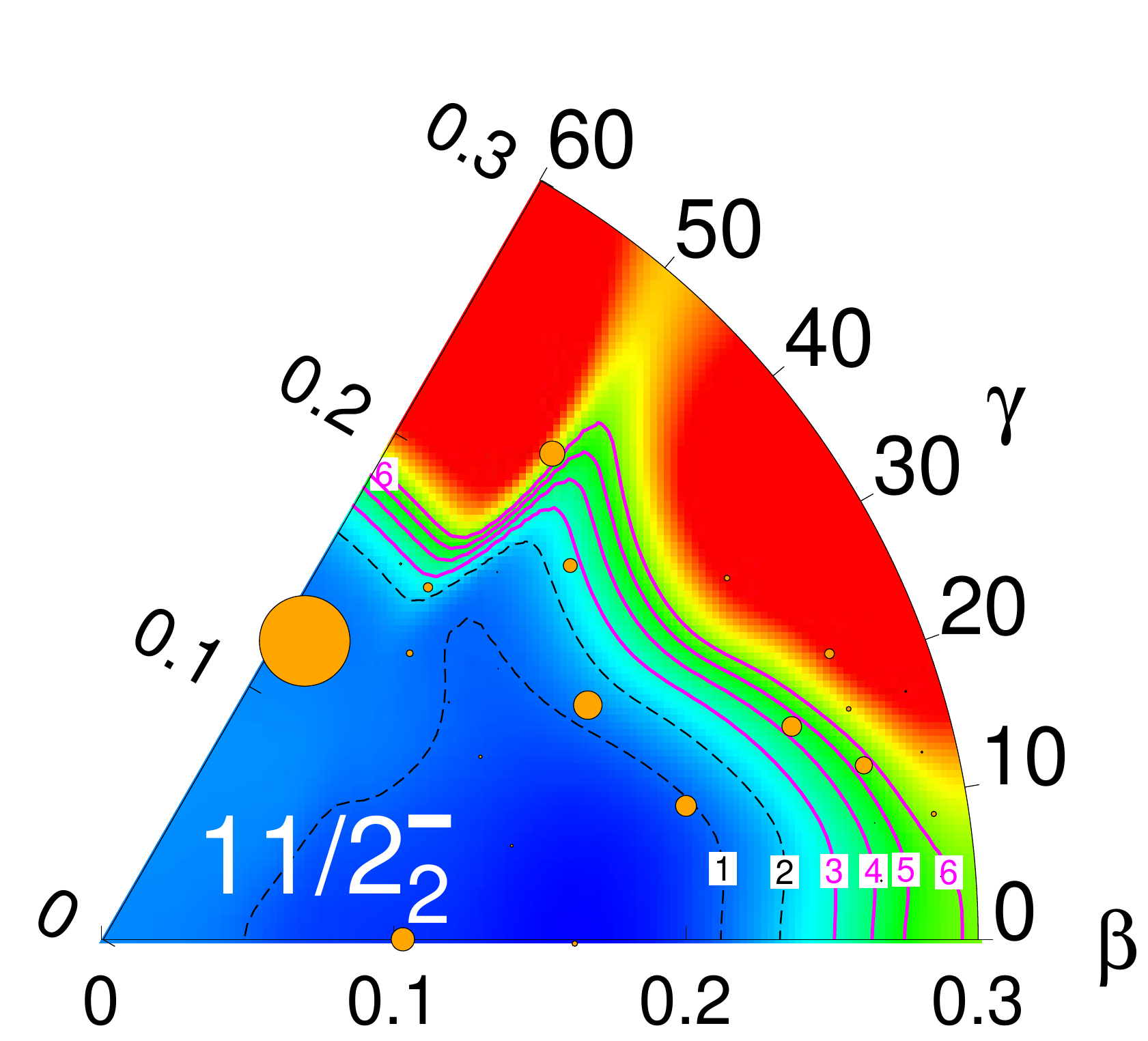}
  \caption{Structure of considered states in deformations $(\beta,\gamma)$ of $^{87}$As. The radius of circles represents the normalized probability to find a deformation $(\beta,\gamma)$ in the corresponding state.}
  \label{beta_gamma_structure_As87}
\end{figure*}

The quadrupole properties  within the valence space can be anticipated with the pseudo-SU3 symmetries as already shown in Ref.~\cite{Ni78-coreII}. This means we adopt the $f_{5/2}, p_{3/2}$ and $p_{1/2}$ proton and $d_{3/2}, d_{5/2}, g_{7/2}, s_{1/2}$ neutron orbitals with degenerate single-particle energies and a pure quadrupole-quadrupole interaction. Following reference~\cite{Nilsson-SU3}, and using the Zuker-Retamosa-Poves (ZRP) diagrams (see figure~\ref{ZRP} and ref~\cite{Nilsson-SU3,PPNP}), we can estimate the configurations and quadrupole moments of the most collective cases $^{85,87}$As. 
On the neutron side, the filling of the pseudo-$pf$ SU3 Nilsson orbitals is non-unique, leading to a largest value of +68.3 and +103.9 e\,fm$^2$ for 2 and 4 particles, respectively. Whereas on the proton side, the filling of the pseudo-$sd$ SU3 Nilsson orbitals rises a maximal value of  +66.4 e\,fm$^2$, but this filling is also non unique, with the last proton being in a $K=\frac{1}{2}$ or $K=\frac{3}{2}$ state. With the incorporation of  the effective charges ($e_p$=1.7 and $e_n$=0.7), the pseudo-SU3 limits for the intrinsic quadrupole moments amount to +160.8 and +187.0 e\,fm$^2$ for $^{85}$As and $^{87}$As, respectively. 
Complementary to LSSM calculations and SU3 estimates, we have performed diagonalizations of the effective Hamiltonian DF2882 in the framework of the Discrete Non-Orthogonal Shell Model (DNO-SM) recently developed in Ref.~\cite{DNOSM}. The model, using mean-field and beyond mean-field techniques, consists in an efficient energy-minimization method proposed by E. Caurier~\cite{Caurier1975} to select relevant deformed Hartree-Fock (HF) states in the potential energy surface represented in a $(\beta,\gamma)$ plane. The diagonalization is then performed after the rotational symmetry restoration using angular momentum projection technique. To treat odd-mass nuclei such as the cases of our interest here, we follow the cranking method as proposed in Ref.~\cite{Kasuya2021}. Figures~\ref{pesAs85},~\ref{pesAs87} present the potential energy surfaces (PES) of $^{85}$As and $^{87}$As,respectively, where both of them show a prolate HF minimum with a strong triaxial softness. These minima correspond to intrinsic quadrupole moments of 135.8 and 148.9 e\,fm$^2$ for $^{85}$As and $^{87}$As, respectively, and are in full agreement with the pseudo-SU3/quasi-SU3 predictions above, exhausting up to 80\% of the SU3 limit in both cases. All the considered states are then optimized through the mixing of deformed projected HF states, with the energy-minimization technique over the whole PES starting from the corresponding HF minimum. The spectra of $^{85}$As and $^{87}$As resulting from these calculations are shown in Figures~\ref{specAs85} and~\ref{specAs87}. Our interest here is to focus on the analysis of the states under study in terms of intrinsic quantities, namely, deformations $(\beta,\gamma)$ and the intrinsic angular momentum (i.e. the total angular momentum projection onto the intrinsic frame axis, denoted by $K$) to see whether they show some band structure with definite $K$-quantum numbers. As shown in Figures~\ref{K_structure_As85} and~\ref{K_structure_As87}, there is a non-negligible mixing of various components in the wave functions. However, their structures with a noticeable dominance of the $|K|=3/2$ component seem to indicate the appearance of a $K^{\pi}=3/2^-$ band in both nuclei. Moreover, Figure~\ref{beta_gamma_structure_As87} depicts the structure of $^{87}$As in the PES, where each yellow circle represents the contribution of a given point $(\beta,\gamma)$. To summarise this part, the overall theoretical description points to clear deformation in rotational regime for $^{85}$As, $^{87}$As. The maximum of deformation is achieved in $^{87}$As ($N$=54), in agreement with the maximum of the triaxial deformation in $^{86}$Ge observed in the germanium isotopes by Lettmann et al.~\cite{Lettmann}.

\section{Conclusion}
In conclusion, medium-spin states in $^{83}$As, $^{85}$As and $^{87}$As have been studied. The level scheme of $^{83}$As and $^{85}$As have been extended by the identification of 12 and 5 new levels, respectively. The partial level scheme of $^{87}$As was established for the first time. The spin and parity assignments were made, based on comparison with the state-of-the-art LSSM calculations, and considering that the reaction used favours population of the yrast states. The new spectroscopic data were also interpreted in terms of pseudo-SU3 symmetries, pointing to the moderate prolate deformation of the $^{85}$As and $^{87}$As ground states. The beyond-mean-field calculations with the novel DNO-SM method describe $^{85}$As and $^{87}$As as deformed nuclei, with prolate deformation in both ground states. The excited states in $^{85}$As show prolate-triaxial nature, while $^{87}$As is predicted to be gamma-soft. This can be compared to the results for $^{84}$Ge and $^{86}$Ge \cite{Lettmann}, where maximal triaxial deformation was found in $^{86}$Ge, one proton below $^{87}$As. \\
The confirmation of the presence of deformation all over the region in Ge, As and Se chains may be an important feature to connect with r-process nucleosynthesis scenarios as these nuclei lie in actual predictions for the $r$-process flow simulations (see  Figure 5 of Ref.\cite{Flow-rprocess}). Moreover, the extension of
such collectivity towards more exotic neutron-rich species should be investigated experimentally in the future in the scope of possible vanishing of the Z=28 shell closure in nickel isotopes beyond N=54 as recently suggested in ~\cite{PPNP}.\\

\section{Acknowledgements}
The authors acknowledge support from the European Union 7th framework through ENSAR, Contract No. 262010. PJ ackowledges support for this work provided by the National Research Foundation (NRF) of South Africa under grant number 90741.\\


\normalsize
\begin{thebibliography}{99}
\bibitem{78Ni} R. Taniuchi et al., Nature, vol. 569 (2019), pp 53-58
\bibitem{Nowacki} F. Nowacki et al., Phys. Rev. Lett. 117, 272501 (2016)
\bibitem{Lettmann} M. Lettmann et al., Phys. Rev. C 96, 011301(R) (2017)
\bibitem{Verney}  M. Lebois et al., Phys. Rev. C 80, 044308 (2009)
\bibitem{Gottardo80Ge} A. Gottardo et al., Phys. Rev. Lett. 116, 182501 (2016)
\bibitem{Triumf80Ge} F. H. Garcia et al.,  Phys.  Rev.  Lett. 125, 17250 (2020)
\bibitem{ISOLDE80Ge} S. Sekal et al., Phys. Rev. C 104, 024317 (2021)  
\bibitem{Theor80GeYES} Da-Li Zhang and Cheng-Fu Mu Chinese Phys. C 42, 034101 (2018)
\bibitem{LSSM-BMF}  K. Sieja et al., Phys. Rev. C 88, 034327 (2013)
\bibitem{Se86}  T. Materna et al., Phys. Rev. C 92, 034305 (2015)
\bibitem{Se-Riken} S. Chen et al., Phys. Rev. C 95, 041302(R) (2017)
\bibitem{Dud81Ga} J. Dudouet et al., Phys.  Rev.  C 100, 011301(R) (2019)
\bibitem{AGATA} S. Akkoyun et al., Nucl.  Instr.  Meth.  A 668, 26 (2012)
\bibitem{Korichi} A. Korichi and T. Lauritsen, Eur. Phys. J.  A 55, 121 (2019)
\bibitem{VAMOS1} M. Rejmund et al., Nucl.  Inst.  and Meth. A 646, 184 (2011)
\bibitem{VAMOS2} M. Vandebrouck et al., Nucl.  Inst.  and Meth. A, 812, 112 (2016)
\bibitem{Clement} E. Clement et al.,  Nucl.  Inst. and Meth. A 855, 1 (2017)
\bibitem{Venturelli} R. Venturelli and D. Bazzacco,  LNL Annual Report 2004,  p. 220 (2005)
\bibitem{track} A. Lopez-Martens et al., Nucl.  Inst.  and Meth.  A 533, 454 (2004)
\bibitem{Dud96Kr} J. Dudouet et al., Phys.  Rev.  Lett 118, 162501 (2017)
\bibitem{Navin}  A. Navin et al., Phys. Lett. B 767, 480 (2017)
\bibitem{83GeBeta} J. A. Winger, et al.,Phys.  Rev.  C 38, 285 (1988)
\bibitem{84GeBeta} A. Korgul et al.,  Phys.  Rev.  C 93, 064324 (2016)
\bibitem{Porquet} M.-G. Porquet et al., Phys. Rev. C 84, 054305 (2011)
\bibitem{Sahin} E. Sahin et al., Nucl. Phys. A 893, 1 (2012)
\bibitem{Drouet} F. Drouet et al., EPJ Web Conf. 62, 01005 (2013)
\bibitem{Baczyk} P. Baczyk et al., Phys. Rev. C 91, 047302 (2015)
\bibitem{Korgul17}  A. Korgul et al.,  Phys. Rev. C 95, 044305 (2017)
\bibitem{Nyako2}B. M. Nyakó et al., Phys. Rev. C 104, 054305 (2021)
\bibitem{Rubidium1} L. Kaubler, et al., Z. Phys. A 352, 127 (1995)
\bibitem{Rubidium3} J. K. Hwang et al.,Phys. Rev. C 80, 037304 (2009)
\bibitem{Ni78-coreI} K. Sieja, F. Nowacki, K. Langanke and G. Martinez-Pinedo, Phys. Rev. C79, 064310 (2009)
\bibitem{Nyako1} B. M. Nyakó et al., Phys. Rev. C 103, 034304 (2021)
\bibitem{NNDC} https://www.nndc.bnl.gov/
\bibitem{Ni78-coreII} K. Sieja, T. R. Rodriguez, K. Kolos and D. Verney, Phys. Rev. C 88, 034327 (2013)
\bibitem{JUN45} M. Honma, T. Otsuka, T. MiZusaki and M. Hjorth-Jensen, Phys. Rev. C 80, 064323 (2009)
\bibitem{GCN5082-I} E. Caurier, J. Menendez, F. Nowacki, A. Poves, Phys. Rev. Lett. 100, 052503 (2008)
\bibitem{GCN5082-II} E. Caurier, F. Nowacki, A. Poves, K. Sieja, Phys. Rev. C 82, 064304 (2010)
\bibitem{Core-pol} K. Sieja and  F. Nowacki, Nuc. Phys. A 857, 9 (2011)
\bibitem{Nilsson-SU3} A. P. Zuker, A. Poves, F. Nowacki and S. M. Lenzi, Phys. Rev. C 92, 024320 (2015).
\bibitem{PPNP} F. Nowacki, A. Obertelli, and A. Poves, Prog. Part. Nuc. Phys. 120, 103866 (2021).
\bibitem{DNOSM} D. D. Dao and F. Nowacki, arXiv:2203.01023, submitted to Phys. Rev. C (2022)
\bibitem{Caurier1975} E. Caurier, Proc. on GCM, BLG report \textbf{484}, 200 (1975).
\bibitem{Kasuya2021} H. Kasuya and K. Yoshida, Prog. Theor. Exp. Phys. 013D01 (2021)
\bibitem{Flow-rprocess} M. P. Reiter et al., Phys. Rev. C 101, 025803 (2020)
\end{thebibliography}
\end{document}